\documentclass[letterpaper,twocolumn,10pt]{article}
\usepackage{usenix2019_v3}

\usepackage{color, colortbl}
\definecolor{Gray}{gray}{0.9}

\usepackage{xspace}
\usepackage{amsmath,amsfonts}
\usepackage{algorithmic}
\usepackage{graphicx}
\usepackage{textcomp}

\def\BibTeX{{\rm B\kern-.05em{\sc i\kern-.025em b}\kern-.08em
    T\kern-.1667em\lower.7ex\hbox{E}\kern-.125emX}}

\usepackage{tikz}
\usepackage{amsmath}

\usepackage{eucal}
\usepackage{filecontents}
\usepackage{epsfig,endnotes}
\usepackage{xspace}
\usepackage{amsmath}

\usepackage{pifont}
\usepackage{todonotes}
\usepackage{hyperref}
\usepackage[normalem]{ulem}
\usepackage{balance}
\usepackage{gensymb}
\usepackage{multirow}
\usepackage{subcaption}
\usepackage{caption}
\usepackage{mathtools}
\usepackage{bbm}
\usepackage{pifont}
\usepackage{balance}
\usepackage{flushend}
\usepackage{booktabs}
\usepackage{makecell}

\newcommand{\mypara}[1]{\medskip\noindent{\bf {#1}.}\xspace}

\newcommand{\etal}{\textit{et al}.}
\newcommand{\eg}{e.g., \xspace}
\newcommand{\ie}{i.e., \xspace}

\usepackage[export]{adjustbox}

\usepackage{cellspace}
\newcommand\cincludegraphics[2][]{\raisebox{-0.3\height}{\includegraphics[#1]{#2}}}

\usepackage[hang,flushmargin]{footmisc}
\usepackage{titlesec}
\titlespacing*{\section}{0pt}{*3}{4pt}
\titlespacing*{\subsection}{0pt}{*2}{3pt}
\titlespacing*{\subsubsection}{0pt}{*1}{2pt}
\let\oldbibliography\thebibliography
\renewcommand{\thebibliography}[1]{%
  \oldbibliography{#1}%
  \setlength{\itemsep}{4pt}%
}

\author{
{\rm Yun Shen}\\
Norton Research Group \\
\and
{\rm Pierre-Antoine Vervier}\\
Norton Research Group \\
\and
{\rm Gianluca Stringhini}\\
Boston University \\
}

\begin{document}
\title{A Large-scale Temporal Measurement of Android Malicious Apps:\\ Persistence, Migration, and Lessons Learned} 

\maketitle

\begin{abstract}
We study the temporal dynamics of potentially harmful apps (PHAs) on Android by leveraging 8.8M daily on-device detections collected among 11.7M customers of a popular mobile security product between 2019 and 2020\let\thefootnote\relax\footnotetext{This paper will appear at the 2022 USENIX Security Symposium}.
We show that the current security model of Android, which limits security products to run as regular apps and prevents them from automatically removing malicious apps opens a significant window of opportunity for attackers.
Such apps warn users about the newly discovered threats, but users do not promptly act on this information, allowing PHAs to persist on their device for an average of 24 days after they are detected.
We also find that while app markets remove PHAs after these become known, there is a significant delay between when PHAs are identified and when they are removed: PHAs persist on Google Play for 77 days on average and 34 days on third party marketplaces.
Finally, we find evidence of PHAs migrating to other marketplaces after being removed on the original one. 
This paper provides an unprecedented view of the Android PHA landscape, showing that current defenses against PHAs on Android are not as effective as commonly thought, and identifying multiple research directions that the security community should pursue, from orchestrating more effective PHA takedowns to devising better alerts for mobile security products.
\end{abstract}

\section{Introduction}
\label{sec:introduction}

Millions of malicious Android apps have been observed over the years~\cite{allix2016androzoo}, performing a variety of malicious activity from sending premium SMS messages~\cite{mirzaei2019andrensemble}, to displaying annoying advertisements~\cite{suarez2020eight}, to enabling stalking~\cite{chatterjee2018spyware}. 
Malicious apps on Android often come in the form of repackaged apps, where a useful Android app is modified to contain hidden malicious functionality to entice users into installing it~\cite{li2017understanding,suarez2020eight,zhou2012detecting}.
To cover the variety of malicious apps that target Android, Google has coined the term \emph{Potentially Harmful Apps} (PHAs).\footnote{\url{https://developers.google.com/android/play-protect/potentially-harmful-applications}}

A large body of research has been published measuring the threat of PHAs on Android.
Previous studies have mostly relied on crawling app markets to retrieve malicious applications~\cite{allix2016androzoo,mirzaei2019andrensemble,suarez2020eight,zhou2012dissecting,wang2018android}.
Alternative approaches include downloading firmware from public repositories to study pre-installed Android apps~\cite{gamba2020analysis} and setting up public analysis infrastructures relying on third parties to submit apps that they suspect are malicious~\cite{lindorfer2014andrubis}.
These approaches then analyze the collected apps by either performing static or dynamic analysis.
While useful to shed light on the functionalities of malicious Android apps, these approaches do not have visibility on the population of infected devices and on how users interact with PHAs.
An alternative approach relied on users installing an app able to monitor network traffic on devices, looking for security and privacy sensitive information~\cite{razaghpanah2018apps}.
This solution solves the aforementioned problem, but it is challenging to recruit a large and representative population of users; in fact, previous studies relied on 11k users to perform their measurements~\cite{razaghpanah2018apps}. 
A third approach that researchers followed is monitoring the network traffic of a mobile ISP and identifying malicious connections based on blacklist information~\cite{lever2013core}.
This approach provides a real-time view of malicious activity from a large number of devices but is limited to monitoring connections to known malicious hosts.
Additionally, this method is limited by the pervasive use of encryption, and does for example allow to observe when a device connects to an app store, but not to inspect what specific PHA a user is installing.

In this paper, we present the first large-scale study to understand the temporal dynamics of PHA installations on Android.
We collect anonymized information about PHA installations from users who installed a popular mobile security product and opted into data collection.
Between 2019 and 2020 we observed over 8.8M PHAs installed on over 11.7M devices from across the globe.
This data allows us to develop a number of metrics and answer the following key research questions:

\begin{table}[t]
\centering

\resizebox{0.95\linewidth}{!} {
\begin{tabular}{llr}    \toprule
\textbf{Dataset} & \textbf{Data} & \textbf{Count}  \\
\midrule

Mobile PHA detection log    & Total records  & 3.2B \\
(01/2019 - 02/2020) & Days           & 416 \\
 & Countries and regions & 201 \\
 & Devices               & 11.7M \\
 & Distinct app names  & 2.3M    \\ 
 & Distinct app SHA2s  & 8.8M  \\ \hline
  
VT                 & Total reports           & 8.8M \\
 & PHA SHA2s (detections $\geq 4$)      & 7M \\
 & Singleton SHA2s (w/o family) & 1.3 M \\
 & PHA SHA2s w/ family & 5.7 M \\
 & PHA families & 3.2K \\ 
\bottomrule
\end{tabular}
}
\caption{Summary of datasets.}
\label{tab:datasets}
\end{table}

\noindent\textbf{How long do devices stay infected with PHAs?} Mobile security products on Android run as regular users without root privileges and are therefore limited in the actions they can take after they detect a malicious program.
Typically, they just raise an alert informing the user about it, and relying on them uninstalling the malicious app.
Our study shows that users do not act promptly on these alerts, and that PHAs persist on devices for approximately 20 days once detected.

\noindent\textbf{How long do PHAs survive on app markets?} By observing millions of mobile devices installing malicious apps from app stores, we can estimate when a certain PHA is removed from the store.
We find that, on average, PHAs persist on Google Play for 77 days, while they persist on alternative marketplaces for 34 days on average.

\noindent\textbf{Do PHAs migrate to other app markets once removed?} We observe 3,553 PHAs that exhibit inter-market migration.
However, those PHAs have on average shorter lifespans in these markets compared to the average persistence time.

\noindent\textbf{Do PHAs persist on devices for longer if migrating from backup/clone services?} 
Android devices allow users to backup their apps and automatically install them on a new device when the user gets a new phone.
We discover that these PHAs on average persist on these devices for longer periods.
For example, we find 14K PHAs that migrated to 35.5K new Samsung devices by using the Samsung smart switch mobile app (\texttt{com.sec.android.easyMover}).
These apps persist in the new devices for 93 days on average.

\noindent\textbf{Implications for Android malware research.} 
Our study has a number of implications for the computer security research community.
We show that malicious apps can survive for long periods of time on app markets, and that the Android security model severely limits what mobile security products can do when detecting a malicious app, allowing PHAs to persist for many days on victim devices.
Furthermore, our results show that the current warning system employed by mobile security programs is not effective in convincing users to promptly uninstall PHAs. This could be due to usability issues such as alert fatigue~\cite{sasse2015scaring}, and calls for more research in this space.
We also show that malicious app developers often move their PHAs to alternative marketplaces after they have been removed.
This suggests that an effective mitigation strategy should include cooperation between multiple marketplaces.

\section{Datasets}
\label{sec:datasets}

\begin{figure}
    \centering
    \includegraphics[width=0.9\linewidth]{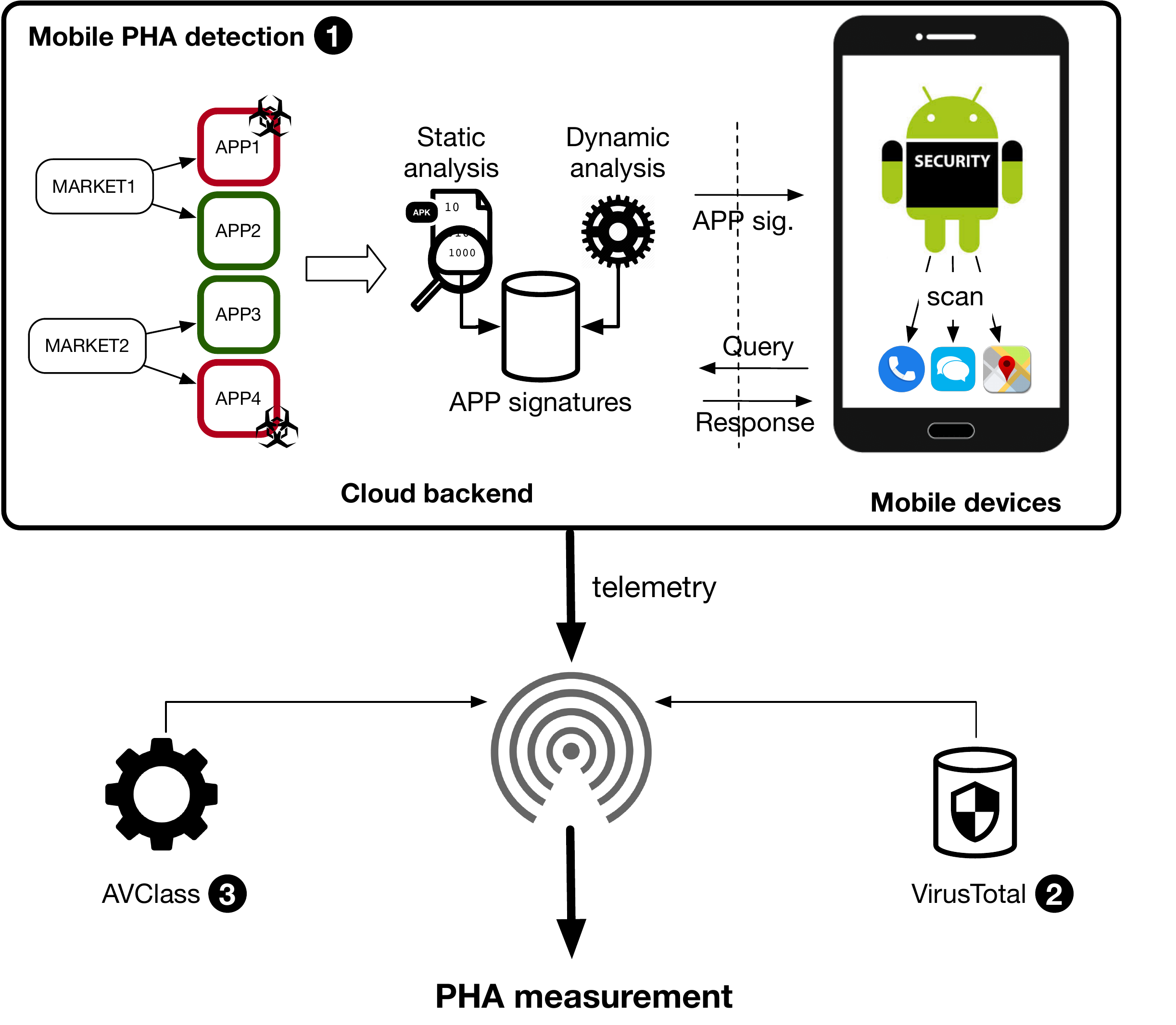}
    \caption{Data Collection Architecture.}
    \label{fig:architecture}
\end{figure}

This section summarizes our data collection approach (see Figure~\ref{fig:architecture}) and the datasets used in this study (see Table~\ref{tab:datasets}).

\mypara{Mobile PHA reputation data \ding{182}} 
In this paper, we use mobile PHA reputation data collected from real-world Android devices by NortonLifeLock's mobile security product, which covers over 200 countries and regions in the world.
Similar to the device geolocation distribution discussed in Kotzias~\etal~\cite{platon2021how}, 25\% of the devices used in our study were from the United States, 28\% of the devices were from European countries, and 31\% of the devices were from the APJ area (although this distribution was skewed towards Japan and India).

NortonLifeLock has an elastic infrastructure to collect and carry out systematic static (\eg flow and context-sensitive taint analysis, fine-grained permission analysis) and dynamic analysis (\eg apk fuzzing, UI-automation, examining network traces, behaviors, etc. in a sandbox environment) of mobile apps from multiple markets and partners at scale.
During the process, nefarious activities and their triggering conditions (such as reflection, dynamic code loading, native code execution, requesting permissions not related to its advertised description, etc.) are analyzed and fingerprinted.
NortonLifeLock employs state-of-the-art commercial products to deal with challenges such as emulator/motion evasion, obfuscated code/libraries, and other evasive techniques, as well as to trigger the critical execution paths in apps. 
The results are then included in NortonLifeLock's detection engine and deployed in its mobile security product to scan and identify suspicious apps on the mobile endpoints.
NortonLifeLock also builds machine learning models from the static and dynamic analysis results of known PHAs and applies these models to inspect unknown or low-prevalence apps.
Also, apps are periodically re-inspected by the analysis infrastructure.

At runtime, the mobile security product periodically scans newly installed apps on a device and can perform a full device scan when requested by the end-user.
When having network access, the security engine queries a cloud backend to obtain the verdicts of the apps installed on a device.
The query contains certain metadata including timestamp, app hash, package name, certificate information, etc. 
The response from the backend includes the reputation scores of the on-device apps together with other proprietary data to guide further actions.
When network access is not available, the security engine leverages the locally stored signatures to scan and identify suspicious apps on the mobile endpoints.
The corresponding scan metadata will then be sent back once network access is restored.

From the backend telemetry data lake, we extract the following information: anonymized device identifier, device country code, detection timestamp, app SHA2, app package name, and installer package name.
This way, we are able to tell the time at which a PHA is detected, on which device it is installed, and which package installed it.
We collected 416 days of detection data between January 1, 2019 and February 20, 2020.
On average, we collect 8M raw events daily (\ie 3.2B events in total).
Note that to carry out the temporal measurement, we only select apps (per SHA2) that we observe at least twice on the same device.
This way, we can reliably calculate their lifespan both on-device and in-market (see Section~\ref{sec:design}).
In total, our dataset covers 2.3M unique package names with 8.8M unique SHA2s from 11.7M devices.
We provide a detailed discussion of bias potentially incurred by our dataset in Section~\ref{sec:discussion}.

\mypara{VirusTotal \ding{183}} 
Note that different security companies have different policies when flagging PHAs (especially adware).
That is, a PHA flagged by NortonLifeLock that collaborated on this study may not have the consensus from other security companies. 
To minimize false positives and bias potentially incurred by our dataset, we query the 8.8M SHA2s corresponding to the PHAs in our dataset on VirusTotal. 
We consider an app as a PHA if VirusTotal returns a minimum of four detections in this paper. 
This is in line with the best practices recently proposed in the malware research community~\cite{zhu2020measuring,platon2021how}.
We refer the audience to Kotzias~\etal~\cite{platon2021how} and Zhu~\etal~\cite{zhu2020measuring} for in-depth analysis of the impact of different detection threshold values of VirusTotal reports.
In total, we identify 7M unique malicious SHA2s, and 3.5K PHA families.

\mypara{AVclass \ding{184}} 
In our study, reliable PHA labeling is a necessary condition to guarantee the quality of malware family attribution.
To this end, we use AVclass~\cite{sebastian2016avclass} to extract family information from AV labels. 
This tool selects the top ranked family corresponding to a majority vote from the VirusTotal report of a given PHA, effectively removing noise in the labels. 
In total, the observed PHAs belong to 3.2K families.
Not all PHAs are equally harmful.
While some apps are clearly malicious (\ie mobile malware including ransomware, Trojans, spyware, etc), others are merely an annoyance to users (\eg adware).
Google groups these apps into Mobile unwanted software (MUwS) as ``apps that are not strictly malware, but are harmful to the software
ecosystem''~\cite{googlereport}.
To investigate differences in how malware and MUwS behave, we use the feature provided by AVclass to classify a sample as Mobile unwanted software (MUwS) or mobile malware (see Section~\ref{sec:pha_device}).
Note that {\sc Euphony}~\cite{hurier2017euphony} also mines AV labels and analyzes the associations between all labels given by different vendors to unify common samples into family groups.
Due to their comparable labeling accuracy in terms of family attribution and the lower memory required by AVclass, we opt for AVclass in this paper.

\mypara{Data distillation and measurement data selection} 
To study the provenance of PHAs, and in particular, which marketplaces they are installed from, we need to collect information on the installer package names of the detected PHAs.
The mobile security product uses the Android API to record a PHA's installer package name when a detection event is triggered. 
However, due to the well known fragmentation from Android device manufacturers and limitations of our measurement infrastructure (\eg we cannot identify an installer package's certificate via Android API), it is hard to accurately extract and attribute the installer packages of all detected PHAs.
For instance, if an app was already installed on a device before the observation period started, our approach would not be able to attribute it to the app that installed it. 
Similarly, if an updated version of an existing PHA was installed, this would be identified as being installed by an update component and not by a marketplace (\eg \texttt{com.google.android.packageinstaller}).
To mitigate this issue, we first identify 3.7M out of 11.7M devices that have at least one PHA installed. 
We then distill the aforementioned datasets by selecting 2.46M devices in which we can attribute their on-device PHAs to the respective installer packages with high confidence. 
In total, we identify 197K PHAs from 2.46M devices that we use in Section~\ref{sec:pha_market} and~\ref{sec:migration} to study the dynamics between PHA, devices, and markets.
These PHAs account for 22\% of all installations recorded by our dataset during the observation period.
We provide a detailed discussion on the limitations of this approach in Section~\ref{sec:discussion}.

\mypara{Ethics and Data Privacy} 
The data used in this paper is privacy sensitive.
NortonLifeLock offers end users the possibility to explicitly opt-in to its data sharing program to help improve the security product's detection capabilities.
This dialog is shown during the setup process when the app is run for the first time, 
and it informs the end-user about the purpose of the telemetry collection, and how the global privacy policy of NortonLifeLock safeguards the data.
For instance, the license agreement specifies that the telemetry ``is processed for the purposes of delivering the product by alerting you to potentially malicious applications, malware, and links'' and ``for the purpose of understanding product usage to further develop and improve the product performance as well as telemetry.''
Since the analysis performed in this paper allows the community to get a better understanding of the Android PHA ecosystem and guide mitigation techniques, this falls under the primary use of the data that users agreed to.
The telemetry data collection, storage, and process are guarded by NortonLifeLock's rigorous privacy policies. 
To preserve the anonymity of users and their devices, client identifiers are anonymized and it is not possible to link the collected data back to the users and the mobile devices that originated it. 
Also, NortonLifeLock does not track the devices or profile user behavior nor has the capability to inspect network data. 
For our measurement study, the anonymized device identifier is only used to compute device-based prevalence rates.  
As such, we are not using any PII and the risks to the users are minimal.

\section{Approach}
\label{sec:approach}

\begin{table}[]
\centering

\resizebox{0.9\linewidth}{!} {
\begin{tabular}{l|l}
\toprule
\textbf{Notation} & \textbf{Description} \\ \midrule
\rowcolor{Gray}
	 $p \in \mathbf{P}$   & a PHA    \\
     $d \in \mathbf{D}$   & a device \\
\rowcolor{Gray}
	 $m \in \mathbf{M}$   & a market \\
	 $f \in \mathbf{F}$   & a PHA family   \\
\rowcolor{Gray}
     $x_{i^{y}}$           &  \begin{tabular}[l]{@{}l@{}} x on/in y, \\ \eg $p_{i^d}$ denotes a PHA $p_i$ detected on device $d$. \end{tabular}            \\
     $(F)$  &  \begin{tabular}[l]{@{}l@{}} first seen timestamp, \\ \eg $p_{i^d}^{(F)}$ denotes first seen timestamp of a PHA $p_i$ on device $d$. \end{tabular} \\
\rowcolor{Gray}
	 $(L)$  &  \begin{tabular}[l]{@{}l@{}} last seen timestamp, \\ \eg $p_{i^m}^{(L)}$ denotes last seen timestamp of a PHA $p_i$ in market $m$. \end{tabular}  \\
	 $\delta_{x^y}$ & \begin{tabular}[l]{@{}l@{}} lifespan of $x$ on/in $y$, \\ \eg $\delta_{p_{i^d}}$ denotes the lifespan of $p_i$ on a device $d$  \end{tabular} \\
\bottomrule
\end{tabular}
}
\caption{Summary of the notations used in this paper. We use lowercase letters to denote an item and bold uppercase letters to denote sets.}
\label{tab:notations}

\end{table}

In this section, we first introduce the overall relationships among PHAs, installer packages, devices, and markets.
We then describe our overall measurement design philosophy and methods together with examples.

\begin{figure}
	\centering
	\includegraphics[width=0.9\linewidth]{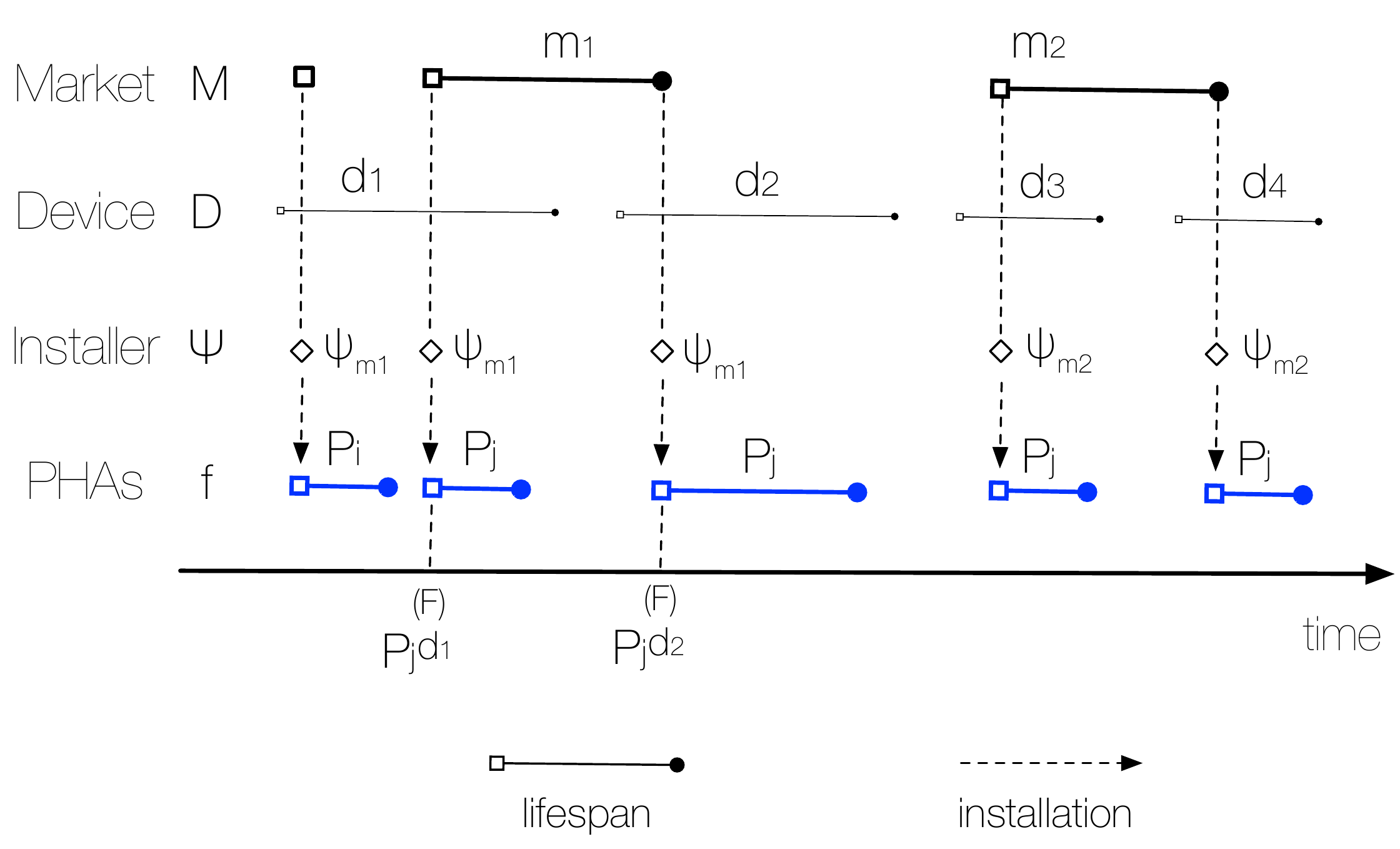}
	\caption{Abstract model of the relations between PHAs, installers, devices, and markets as observed in our dataset.}
	\label{fig:relations}
\end{figure}

\subsection{Relationships} 

For the reader's convenience, we summarize the notations introduced here and in the following sections in Table~\ref{tab:notations}.
We provide a detailed description of the relations observed in our dataset to form the foundation of our measurements in the rest of the paper.
Figure~\ref{fig:relations} shows an example to illustrate the complex dynamic relations among PHAs $\mathbf{P}$, installer packages $\Psi$, devices $\mathbf{D}$, and markets $\mathbf{M}$, coupled with a timeline.
Each device $d$ can have multiple PHAs installed (\eg $d_1$ has two PHAs $p_i$ and $p_j$ in Figure~\ref{fig:relations}).
A PHA $p_j$ can be present in multiple devices (\eg $p_j$ is installed in all four devices).
Additionally, multiple PHAs can belong to a PHA family.
For example, as we can see in Figure~\ref{fig:relations}, $\mathbf{P}_f$ includes $p_i$ in $d_1$ and $p_j$ in all four devices.
In addition, the Android API allows the mobile security product to retrieve the package name (\ie $\psi$) of the application that installed a PHA.
This enables us to identify which market a PHA came from if the package name of $\psi$ matches the name of the market.
For example, $p_j$ on device $d_1$ is installed by a package $\psi_{m_1}$ from market $m_1$ at timestamp $p_{j^{d_1}}^{(F)}$ (see Figure~\ref{fig:relations}).
Aggregating all installation events of the \emph{same} PHA $p_i$ in all devices $\mathbf{D}$, we can estimate the lifespan $\delta_{p_{i^m}}$ in market $m$ as [$p_{j^{d_1}}^{(F)}$, $p_{j^{d_2}}^{(F)}$] (see Figure~\ref{fig:relations}).

\subsection{Design Philosophy}
\label{sec:design}

Measuring the in-market presence of PHAs (\eg how fast PHAs are removed) is a challenging task as we are not the app market owners.
One solution is to crawl known app markets and track all apps on a daily basis~\cite{wang2018beyond}.
However, crawling results cannot be correlated with the device installation data since not all markets offer precise device installation information.
In this study, we adopt an outside-in design philosophy to perform our market presence measurements.
That is, we treat mobile devices as \emph{sensors} and their PHA installation events as the \emph{probing results} of a PHA's existence.
We then use the information on the installer packages of apps to identify the origin markets of installed PHAs (see the above section for relations).
By correlating this information with on-device detection timestamps we can calculate PHA in-market persistence and prevalence in a non-intrusive, outside-in way.
Similarly, we can also calculate PHA on-device persistence using the detection timestamps. 
In this study we use different metrics to study the PHA ecosystem along three axes: on-device persistence, in-market persistence, and PHA migration across markets.
In this section, we define the metrics that we will later use to measure these three aspects.

\subsubsection{Measurement of PHA On-device Persistence}
\label{sec:method_ondevice_persistence}

The mobile security product runs periodically in the background and sends telemetry data to the backend if PHAs are detected.
If a PHA was not removed from the device after the user was displayed an alert, the mobile security product records this recurrent detection at different timestamps until the PHA is removed from the device.
Given this series of detection events, we are able to tell the first seen and last seen timestamps of a PHA $p_i$ on a device $d$, consequently enabling us to estimate the lifespan of $p_i$ on a device $d$ (\ie $\delta_{p_{i^d}}$).
Following this observation, we use Eq~\ref{eq:pha_family_single_device} to measure the persistence period a PHA family $f$ on a device $d$.

\begin{equation}
persistence(f, d) = \sum_{p_i \in \mathbf{P}_f}(\delta_{p_{i^d}})/|\mathbf{P}_f|
\label{eq:pha_family_single_device}
\end{equation}

\noindent That is, we calculate the mean lifespan of all PHAs belonging to a family $f$ on device $d$.
For example, in Figure~\ref{fig:relations}, family $f$ has two PHAs ($p_i$ and $p_j$) on device $d_1$, hence $persistence(f, d_1)$ = ($\delta_{p_{i^{d_1}}}$  + $\delta_{p_{j^{d_1}}}$)/2.
We then use Eq~\ref{eq:pha_family_all_devices} to measure the mean persistence period per PHA family $f$ on all devices $\mathbf{D}$.

\begin{equation}
persistence(f, \mathbf{D}) = \sum_{d \in \mathbf{D}} persistence(f, d)/ |\mathbf{D}|
\label{eq:pha_family_all_devices}
\end{equation}

\noindent For example, family $f$ has presence in all four devices in Figure~\ref{fig:relations}.
Following Eq~\ref{eq:pha_family_all_devices}, we can calculate $persistence(f, \mathbf{D})$ as [$persistence(f, d_1)$ + $persistence(f, d_2)$ + $persistence(f, d_3)$ + $persistence(f, d_4)$]/4.

\subsubsection{Measurement of PHA In-market Persistence}

Given a single device $d$, when the mobile security product detects a PHA on the mobile device, it also records the installer package name of this PHA.
Correlating this with the official package names of the markets, we can identify if a PHA was installed from a certain market $m$ at a certain timestamp.
For example, if we observe the installer package name of a PHA is \texttt{com.android.vending}, we can tell that this PHA comes from the Google Play store.
Note that malicious apps can impersonate the legitimate apps on Android devices (\eg \texttt{com.android.vending} may not be the legitimate Google Play app).
To avoid false attributions, we check the detection telemetry data of the same device and verify if any detection records match the same package names of the known marketplaces.
By doing so, we are able to verify the legitimacy of the market apps in this measurement study.
We provide a detailed discussion of the limitations of this approach in Section~\ref{sec:discussion}. Note that first seen timestamp of a PHA on device $d$ can reliably prove that a PHA exists in a market at the time of installation. By aggregating the first detection events of a PHA $p_i$ across all devices $\mathbf{D}$, we can represent a PHA's in-market appearances using Eq~\ref{eq:pha_market_lifespan}.

\begin{equation}\label{eq:pha_market_lifespan}
	   \Omega_{p_{i^m}} = \{ p_{i^{d_j}}^{(F)} \}, \forall d_j \in \mathbf{D}, p_{i^{d_j}} \in \mathbf{P}_{i^m}
\end{equation}

\noindent Essentially, $\Omega_{p_{i^m}}$ represents a series of timestamps where $p_i$ was first seen on all devices $D$.
Take the relations in Figure~\ref{fig:relations} as an example, we have two detections of a PHA $p_j$ respectively on $d_1$ and $d_2$ installed from market $m_1$.
In turn, we have $\Omega_{p_{j^{m_1}}}$ = $\{ p_{j^{d_1}}^{(F)}, p_{j^{d_2}}^{(F)} \}$.
\noindent Following the above observation, we use Eq~\ref{eq:pha_family_single_sha2} to measure the persistence period of a PHA $p_i$ in a market $m$.

\begin{equation}\label{eq:pha_family_single_sha2}
	   persistence(p_i, m) = max(\Omega_{p_{i^m}}) - min(\Omega_{p_{i^m}})
\end{equation}

\noindent It is straightforward to observe $persistence(p_j, m_1) = p_{j^{d_2}}^{(F)} - p_{j^{d_1}}^{(F)}$ following Eq~\ref{eq:pha_market_lifespan} and Eq~\ref{eq:pha_family_single_sha2}. Note that we rely the on-device detection to measure a PHA's in-market persistence. It is possible that a PHA still exists in a market but our dataset did not reflect its existence. Consequently, we measure the \emph{lower bound} of the PHA in-market persistence. Finally, we use Eq~\ref{eq:pha_family_all_sha2s} to measure the persistence period a PHA family $f$ in a market $m$.

\begin{equation}\label{eq:pha_family_all_sha2s}
 	   persistence(f, m) = \sum_{p_i \in \mathbf{P}_f} persistence(p_i, m) / |\mathbf{P}_f|
\end{equation}

\subsubsection{Measurement of PHA Inter-market Migration}

Recall that the mobile security product records that a PHA $p$ was installed on a device $d$ at a timestamp $t$ by an installer package $\psi$.
By aggregating the telemetry data about a specific PHA $p$ and mapping its installer package names to marketplaces across all devices $\mathbf{D}$, we can track the appearance of a PHA $p_i$ across all marketplaces $\mathbf{M}$.
Take PHA $p_j$ in Figure~\ref{fig:relations} for example, it was detected in four devices ($d_1$, $d_2$, $d_3$, and $d_4$) from two marketplaces ($m_1$ and $m_2$).
Following Eq~\ref{eq:pha_market_lifespan}, the lifespan of $p_j$ in $m_1$ and $m_2$ are respectively $\delta_{{p_j}^{m_1}}$ = $[min(\Omega_{p_{j^{m_1}}}), max(\Omega_{p_{j^{m_1}}})]$ and $\delta_{{p_j}^{m_2}}$ = $[min(\Omega_{p_{j^{m_2}}}), max(\Omega_{p_{j^{m_2}}})]$.
As we observe in Figure~\ref{fig:relations} that $max(\Omega_{p_{j^{m_1}}})$ is less than $min(\Omega_{p_{j^{m_2}}})$, we define that a PHA $p_j$ exhibits \emph{inter-market migration} from $m_1$ to $m_2$.
Following the observation, we use Eq~\ref{eq:pha_appearance} to represent the appearances of a PHA $p_i$ across the marketplaces $\mathbf{M}$.

\begin{equation}\label{eq:pha_appearance}
	   appearance(p_i, \mathbf{M}) = \{\delta_{{p_i}^{m}}\}, \forall m \in \mathbf{M}
\end{equation}

\noindent Note that each $\delta_{{p_i}^{m}}$ is an interval (\ie $[min(\Omega_{p_{i^{m}}}), max(\Omega_{p_{i^{m}}})]$). In turn, we sort $appearance(p_i, \mathbf{M})$ by $min(\delta_{{p_i}^{m}})$, then identify sequentially non-overlapping intervals from $appearance(p_i, \mathbf{M})$ to measure PHA inter-market migration across the marketplaces $\mathbf{M}$.

\subsection{Right Censored Data}
\label{sec:censored}
Censoring occurs when incomplete information is available about the survival time of some individuals. 
Recall that our observation period is between January 1, 2019 and February 20, 2020.
There exist a number of PHAs that we cannot observe if they have been removed from the markets after our study ends on February 20, 2020.
Such PHA data is defined as right censored in survival analysis~\cite{kleinbaum2010survival}.
In our study, we assume that censoring is independent or unrelated to the likelihood of developing the event of interest (\ie PHA removal).
We, therefore, keep these right censored data to avoid estimation bias.
More details can be found in Section~\ref{sec:pha_market}.

\section{Temporal Characteristics of PHA Families}
\label{sec:pha_prevalence}

\begin{table}[t]
\centering
\resizebox{0.95\linewidth}{!} {
\begin{tabular}{ccccc}
\toprule
\textbf{Rank} & \textbf{Family}  &  \textbf{Total SHA2s} & \textbf{\begin{tabular}[c]{@{}c@{}}Active SHA2s/Month \\ (01/19 - 02/20)\end{tabular}} & \textbf{\begin{tabular}[c]{@{}c@{}}\# month \\ $\geq$ avg \end{tabular}} \\ \midrule
\rowcolor{Gray}
1  & jiagu (U)    & 671K & \cincludegraphics[width=1.5in]{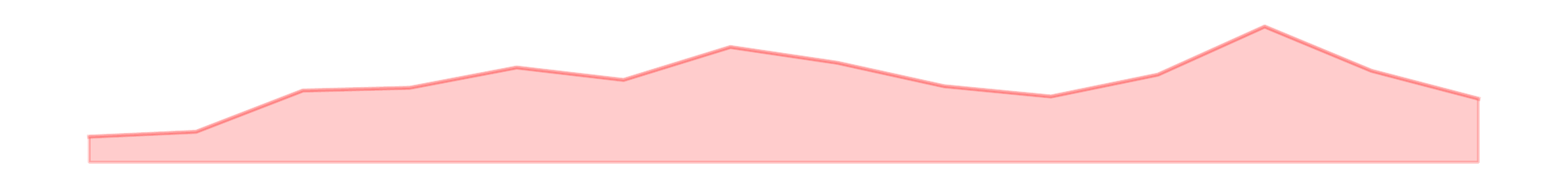} & 7  \\ 
2  & smsreg (M)  & 438K & \cincludegraphics[width=1.5in]{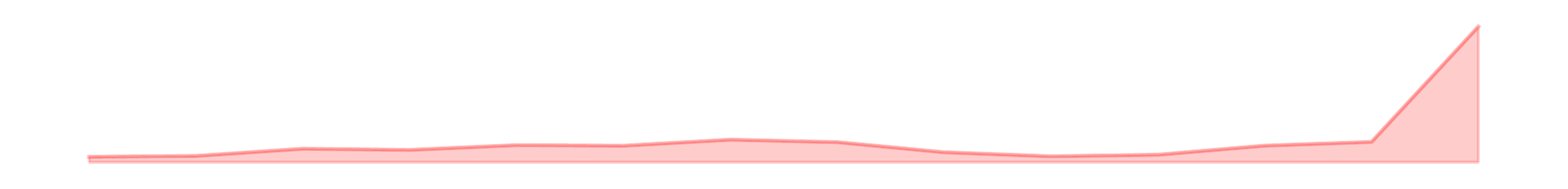} & 2  \\ 
\rowcolor{Gray}
3  & hiddad (U)   & 308K & \cincludegraphics[width=1.5in]{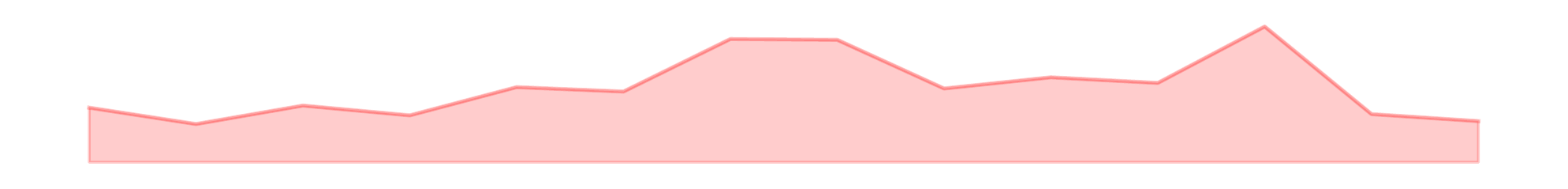} & 6 \\ 
4 & airpush (U)   &  164K & \cincludegraphics[width=1.5in]{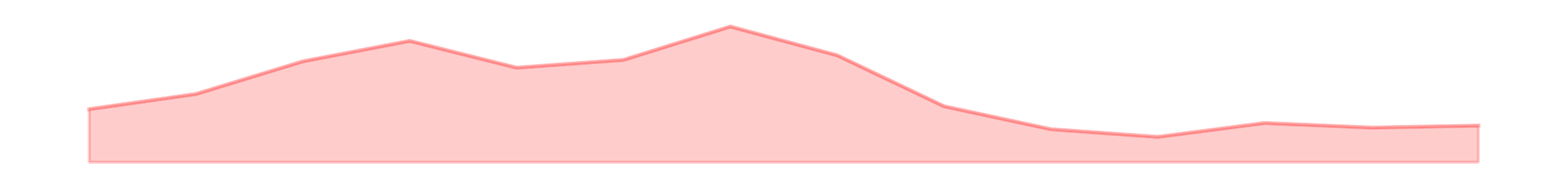}   & 6 \\ 
\rowcolor{Gray}
5  & revmob (U) & 132K & \cincludegraphics[width=1.5in]{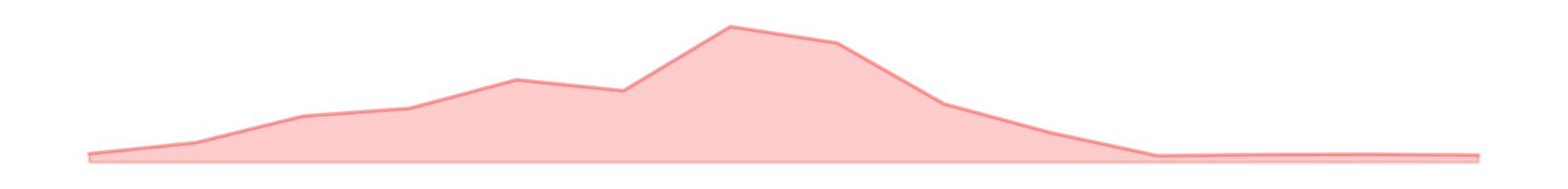} & 6  \\ 
6  & dnotua (U)     &  105K & \cincludegraphics[width=1.5in]{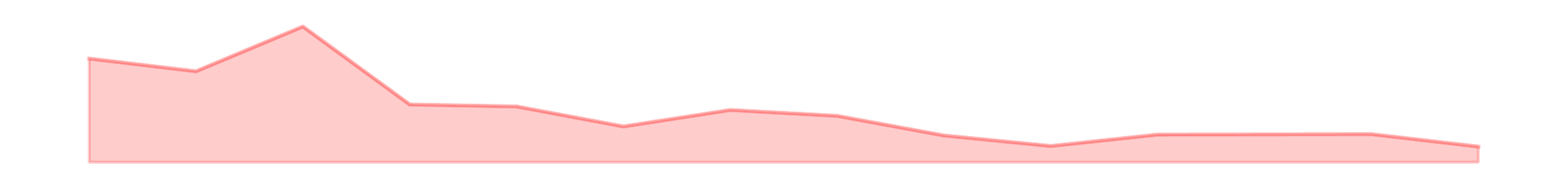} & 6 \\ 
\rowcolor{Gray}
7  & dowgin  (U)   &  87K & \cincludegraphics[width=1.5in]{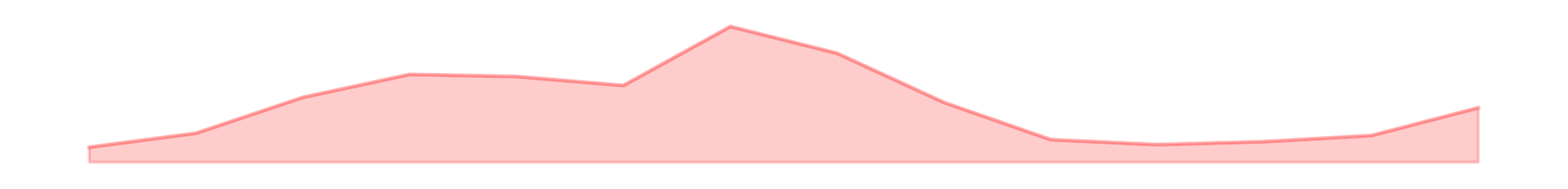}  & 6  \\ 
8  & leadbolt (U)   &  75K & \cincludegraphics[width=1.5in]{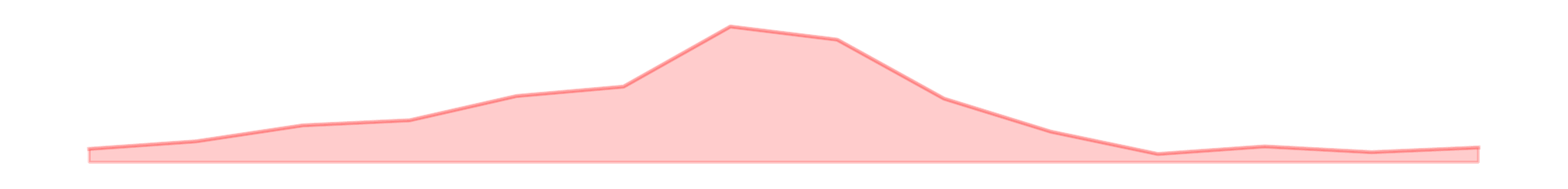} & 7 \\ 
\rowcolor{Gray}
9  & mobidash (U)   & 74K  & \cincludegraphics[width=1.5in]{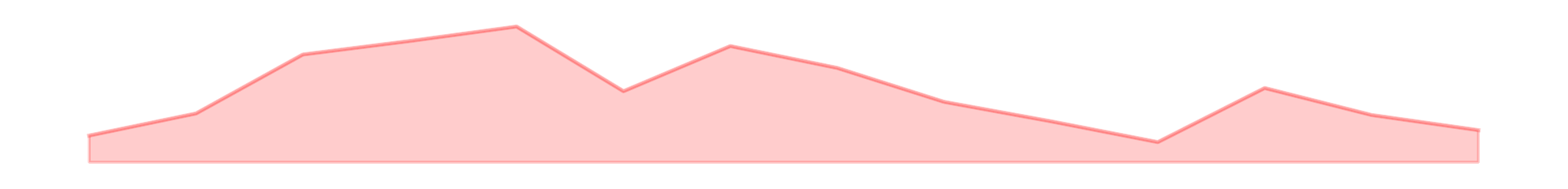}& 5  \\ 
10 & kuguo  (U)   &  72K & \cincludegraphics[width=1.5in]{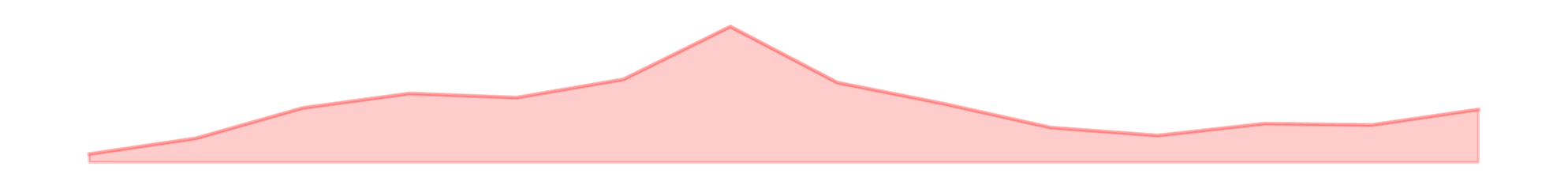}  & 6  \\ 
\rowcolor{Gray}
11 & locker (M)    &  60K  & \cincludegraphics[width=1.5in]{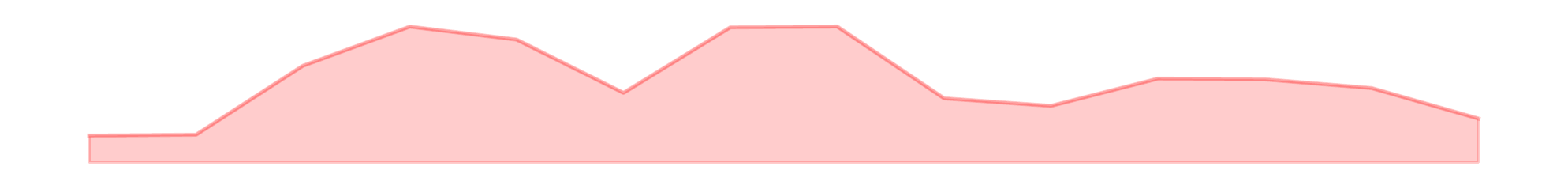} & 6  \\ 
12 & ewind (M)  &  57K & \cincludegraphics[width=1.5in]{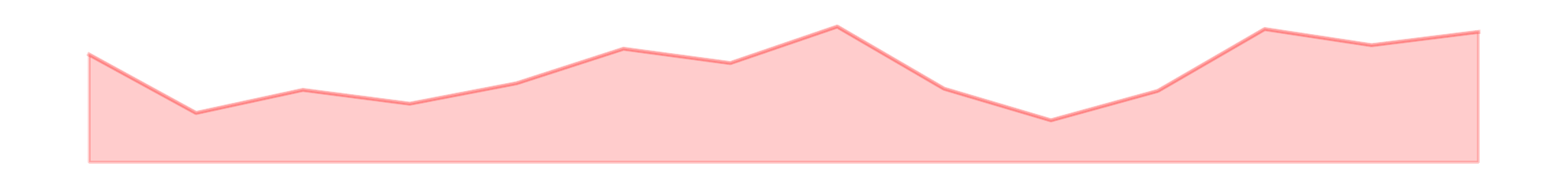} & 7 \\ 
\rowcolor{Gray}
13 & secapk (U)  &  51K  & \cincludegraphics[width=1.5in]{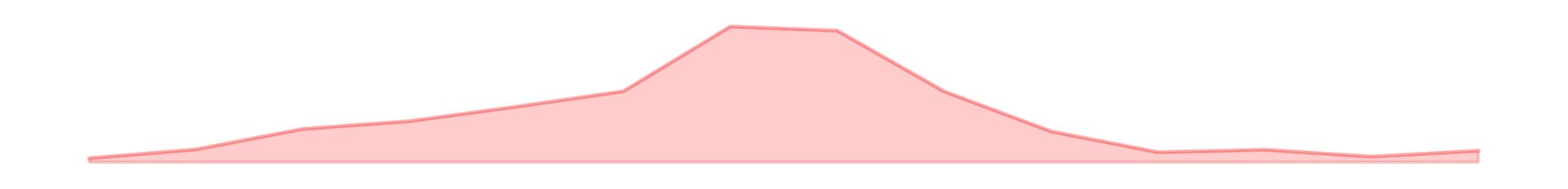}  & 7  \\ 
14  & inmobi (U)   &  44K & \cincludegraphics[width=1.5in]{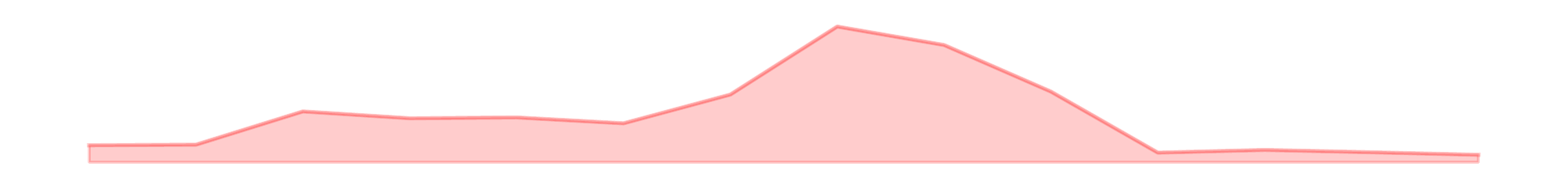}  & 5  \\ 
\rowcolor{Gray}
15 & tencentprotect   &  44K & \cincludegraphics[width=1.5in]{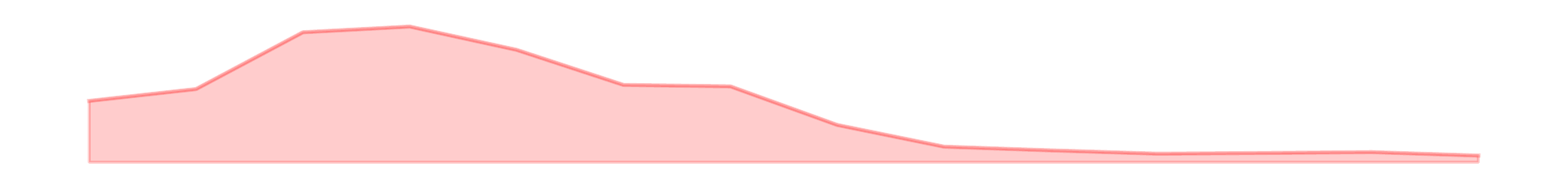}  & 5 \\ 
16 & koler  (M)   &  42K &  \cincludegraphics[width=1.5in]{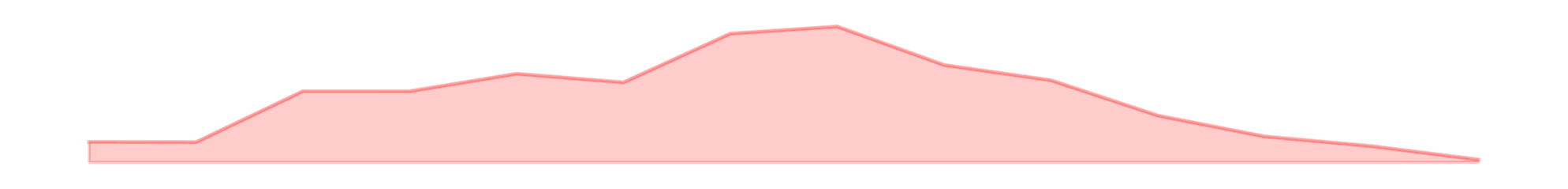}  & 7 \\
\rowcolor{Gray}
17 & domob  (U)  &  40K & \cincludegraphics[width=1.5in]{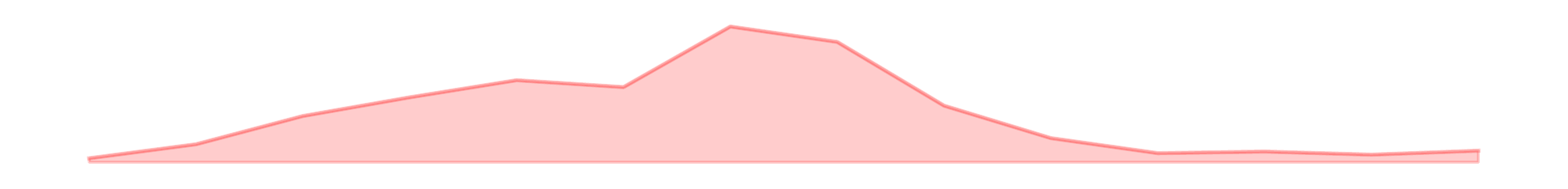}   & 8 \\ 
18 & secneo  (U)  &  29K  & \cincludegraphics[width=1.5in]{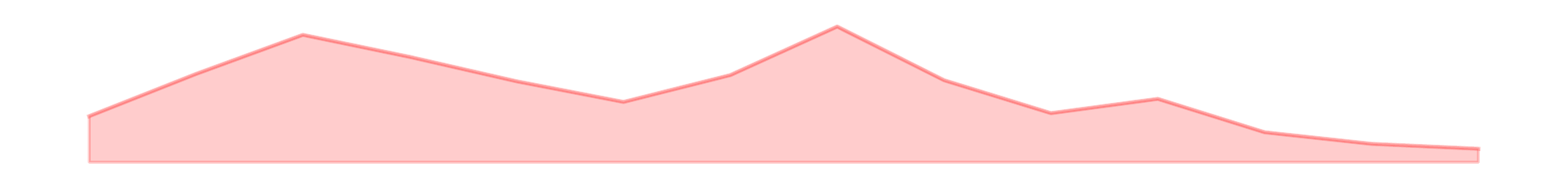} & 7   \\ 
\rowcolor{Gray}
19 & autoins (M)   &  25K & \cincludegraphics[width=1.5in]{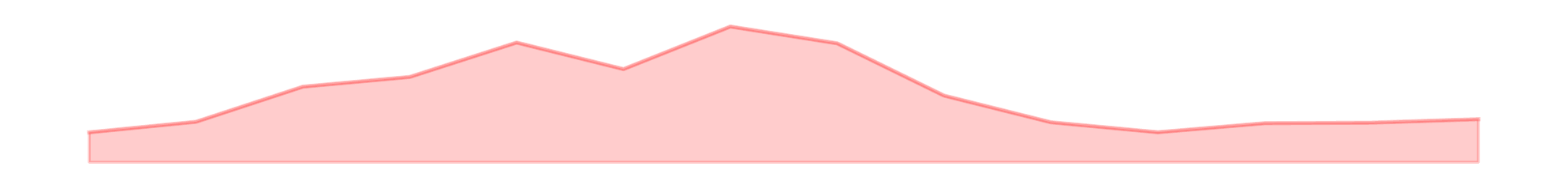} & 6    \\ 
20 & datacollector  (M)  &  15K & \cincludegraphics[width=1.5in]{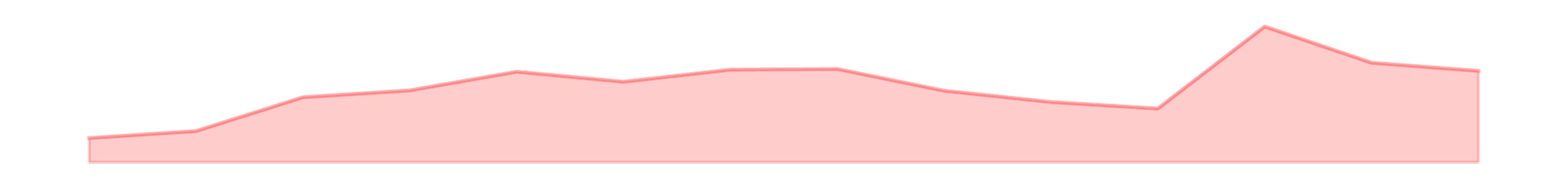}   & 7   \\ 
\bottomrule
\end{tabular}
}
\caption{Summary of the temporal patterns of the top 20 PHA families. These families are ordered by the total number of SHA2s. \textbf{M} denotes Malware and \textbf{U} denotes MUwS/Adware.}
\label{tab:pha_temporal_top_20}
\end{table}

Miscreants either repackage multiple apps with the same malicious code or modify their code to avoid being detected by security measures~\cite{mirzaei2019andrensemble,zhou2012detecting}.
These related malicious apps are commonly referred to as \emph{families}.
In this section, we study the temporal patterns of PHA families (\eg do we see the same top families all the time or do they gradually reduce their operation due to increased detection efforts from mobile security companies?). 
As discussed in Section~\ref{sec:datasets}, we use AVclass~\cite{sebastian2016avclass} to group apps into families.
Table~\ref{tab:pha_temporal_top_20} shows the temporal prevalence of the top 20 PHA families and summarizes our findings.
These top 20 PHA families are detected in 2.01 million devices, approximately 67\% of the aforementioned 3.7M devices.
Recall that we select PHAs (per SHA2) that we observe at least twice on the same device to carry out the measurements (see Section~\ref{sec:datasets}), hence our dataset is purposely designed to measure the temporal behavior.
Our results are consistent with the most recent Android PHA device prevalence study~\cite{platon2021how}: 15 of the 20 PHA families in Table~\ref{tab:pha_temporal_top_20} were also among the top 20 PHA families found by~\cite{platon2021how}. 
As we can see in Table~\ref{tab:pha_temporal_top_20}, 16 out of the 20 top PHA families manage to exceed their average number of monthly active SHA2s for at least 6 months.  
Also, we see that majority of the PHA families exhibit bell shaped temporal patterns.
This shows that these PHA families may eventually reduce their operation due to increased detection efforts from mobile security companies and marketplaces.
We will further analyze this finding in Section~\ref{sec:pha_market} (\eg how rapid takedown can disrupt PHA operations) and, consequently, how miscreants may move their PHAs to alternative markets in Section~\ref{sec:migration}. 
Besides, we notice that \texttt{smsreg} and \texttt{smspay} show an upward pattern towards the end of our observation period (\ie January and February 2020).
In light of the recent discussion of the limitation of SMS-based 2FA authentication\footnote{ https://krebsonsecurity.com/2018/08/reddit-breach-highlights-limits-of- sms-based-authentication/}, our findings indicate that the possibility of such breaches still exists in the wild and has attracted the attention of cybercriminals.

\section{PHA On-Device Persistence}
\label{sec:pha_device}

As we explained, mobile security products are limited by the Android security model, and they lack the ability to delete PHAs once they detect them.
Instead, they typically inform the user about the newly discovered threat, asking them to manually remove the app. 
This leaves the question of how promptly users remove identified PHAs from their devices.
In this section, we first study the PHA on-device persistence to understand how long PHAs can persist on devices once installed. 
We then study the consequences of PHA on-device persistence and, for example, whether this leads to  additional PHA installations on devices.

\subsection{On-Device Persistence of Different PHA Types} 

\begin{table}[t]
\centering
\resizebox{0.9\linewidth}{!} {
\begin{tabular}{cccccc}
\toprule
\multicolumn{2}{c}{\textbf{Overall}} &
  \multicolumn{2}{c}{\textbf{Malware}} &
  \multicolumn{2}{c}{\textbf{MUwS}} \\ \cmidrule(lr){1-2} \cmidrule(lr){3-4} \cmidrule(lr){5-6}
\textbf{\# Devices} &
  \textbf{\begin{tabular}[c]{@{}c@{}}Avg.\\ Persistence \end{tabular}} &
  \textbf{\# Devices} &
  \textbf{\begin{tabular}[c]{@{}c@{}}Avg.\\ Persistence\end{tabular}} &
  \textbf{\# Devices} &
  \textbf{\begin{tabular}[c]{@{}c@{}}Avg.\\ Persistence\end{tabular}} \\ \cmidrule(lr){1-2} \cmidrule(lr){3-4} \cmidrule(lr){5-6}
3.7M &
  20.2 D &
  2.93M &
  20.3 D &
  1.97M &
  13.1 D \\ 
\bottomrule
\end{tabular}
}
\caption{Overall PHA on-device persistence.}
\label{tab:pha_persistence_overall}
\end{table}

As discussed, not all Android PHAs are equally harmful, but some are merely annoying to users (MUwS). 
It is therefore possible that users will react differently when the security product informs them that they have installed malware compared to mobile unwanted software (MUwS), possibly not uninstalling the latter.
To better understand this, we use AVClass~\cite{sebastian2016avclass} to distinguish mobile malware from MUwS among PHAs. 
We then follow the approach outlined in Section~\ref{sec:design} and measure the overall PHA on-device persistence. 
Our findings are summarized in Table~\ref{tab:pha_persistence_overall}.
We find that PHAs persist on devices for approximately 20 days once installed. 
On average, mobile malware can persist longer than MUwS (respectively 20.3 days and 13.1 days). 
It is surprising that end users do not promptly remove PHAs once detected.
The prolonged persistence of PHAs on devices leaves a window of opportunity during which attackers can cause harm to the victims and their devices (\eg displaying intrusive full screen ads, collect private information, install additional malicious apps without user consent).

\subsection{PHA Family On-Device Persistence}

\begin{table}
\centering
\resizebox{0.9\linewidth}{!} {
\begin{tabular}{crrcc}
\toprule
  \textbf{Family} &
  \textbf{\begin{tabular}[c]{@{}c@{}}Avg. \\ Persistence \end{tabular}} &
  \textbf{\begin{tabular}[c]{@{}c@{}}Max\\ Persistence\end{tabular}} &
  \textbf{\begin{tabular}[c]{@{}c@{}}\# Devices\\ ($\leq$ Avg.)\end{tabular}} &
  \textbf{\begin{tabular}[c]{@{}l@{}}\# Devices \\ ($>$ Avg.)\end{tabular}} \\ \midrule
\rowcolor{Gray}
jiagu & 4.77 D & 414.0 D & 1132K  &  303K \\ 
smsreg & 2.37 D & 413.63 D & 471K  &  34K \\ 
\rowcolor{Gray}
hiddad & 5.54 D & 415.08 D & 611K  &  85K \\ 
airpush & 20.9 D & 413.5 D & 183K  &  35K \\ 
\rowcolor{Gray}
revmob & 3.54 D & 413.82 D & 354K  &  13K \\ 
dnotua & 2.93 D & 414.36 D & 261K  &  12K \\ 
\rowcolor{Gray}
dowgin & 7.24 H & 412.11 D & 239K  &  1K \\ 
leadbolt & 5.74 D & 413.45 D & 243K  &  12K \\ 
\rowcolor{Gray}
mobidash & 1.8 D & 415.09 D & 294K  &  10K \\ 
kuguo & 6.18 H & 408.89 D & 155K  &  1K \\ 
\rowcolor{Gray}
locker & 4.13 H & 413.13 D & 195K  &  322 \\ 
ewind & 8.72 D & 414.11 D & 118K  &  22K  \\ 
\rowcolor{Gray}
secapk & 12.98 H & 412.03 D & 215K  &  1K \\ 
inmobi & 10.66 D & 413.77 D & 213K  &  23K \\ 
\rowcolor{Gray}
tencentprotect & 5.27 D & 413.58 D & 173K  &  23K \\ 
koler & 0.49 H & 360.01 D & 160K  &  1K \\ 
\rowcolor{Gray}
domob & 1.87 D & 409.91 D & 174K  &  2K \\ 
secneo & 3.32 D & 413.15 D & 157K  &  11K \\ 
\rowcolor{Gray}
autoins & 22.45 D & 414.0 D & 174K  &  43K \\ 
datacollector & 15.47 D & 413.59 D & 176K  &  54K \\ 
\bottomrule
\end{tabular}
}
\caption{Summary of the top 20 PHA family on-device persistence. \textbf{D} denotes days and \textbf{H} denotes hours.}
\label{tab:pha_device_persistence_top_20}
\end{table}

We then follow the approach outlined in Section~\ref{sec:design} to understand the on-device persistence of the top 20 largest PHA families ranked by their device prevalence ratios. 
Our findings are summarized in Table~\ref{tab:pha_device_persistence_top_20}. 
It is interesting to observe that 15 out of the 20 top PHA families can persist on devices for several days, and only five PHA families are removed by end users in less than two days. 
For example, \texttt{ewind}, a Trojan family, persisted on 118K devices for up to 8.72 days on average and on 22K devices for even longer. 
This is interesting because it indicates that users choose not to uninstall the malicious app after they are warned by the mobile security product.
In light of this, It is interesting to observe that end users removed \texttt{locker/koler} rapidly after detection. 
We can only speculate that the reason behind this might be the degradation in user experience (\ie screen lockdown by \texttt{locker}, fake FBI warnings by \texttt{koler}). 
We hope that our findings would enable mobile security companies to devise effective notification systems to nudge the end users to remove PHAs upon detection, taking for example into account alert fatigue~\cite{sasse2015scaring}.

\subsection{PHA Multiple-Instance Persistence}

\begin{figure}
	\centering
	\includegraphics[width=0.7\linewidth]{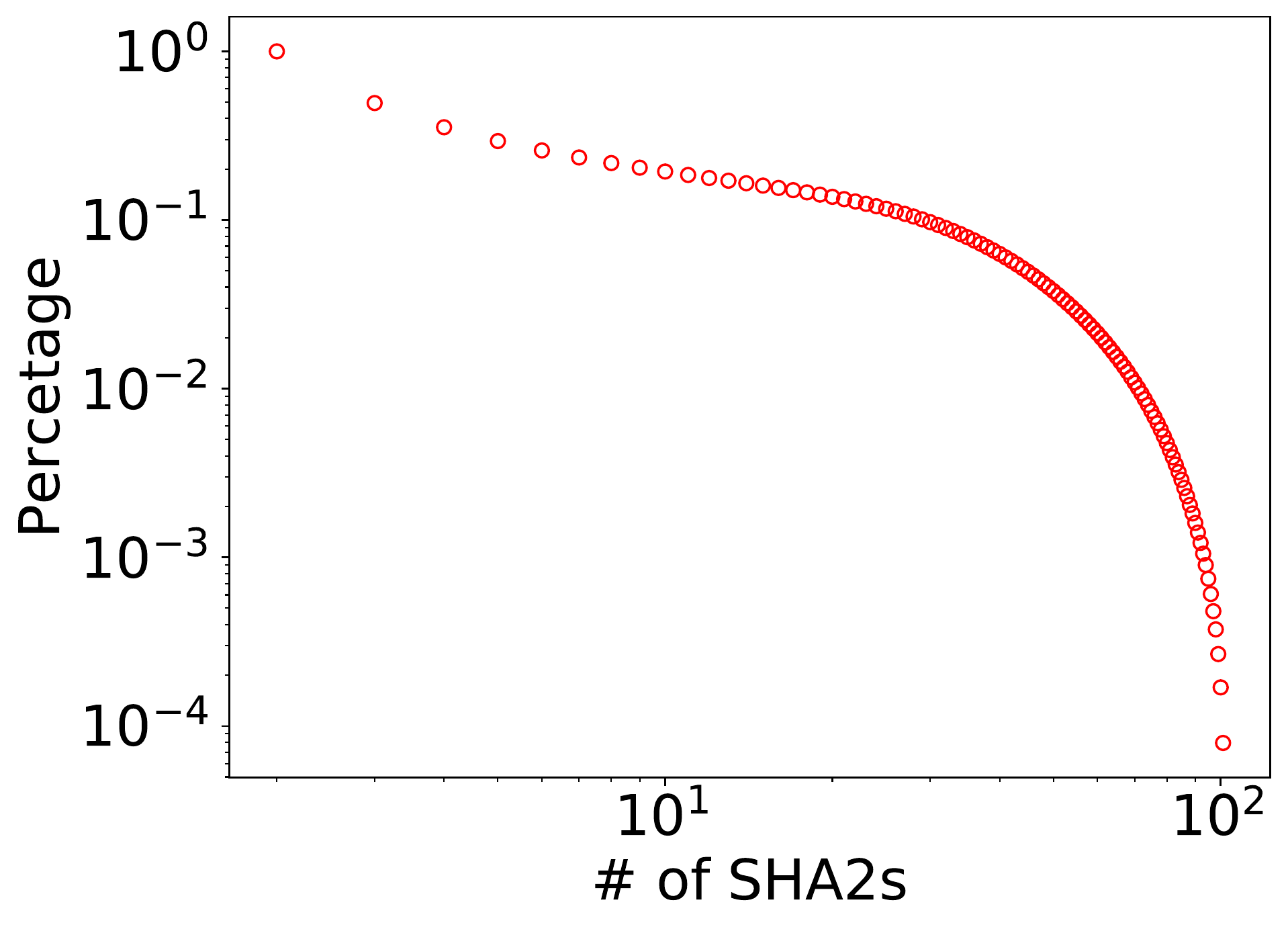}
	\caption{CCDF of the number of SHA2s on devices in our dataset (log scale).}
	\label{fig:PHA_CDF}
\end{figure}

\begin{table}[t]
\centering
\resizebox{0.8\linewidth}{!} {
\begin{tabular}{cccc}
\hline
\textbf{Family} & \textbf{\# Devices} & \textbf{\begin{tabular}[c]{@{}c@{}}\# Multi-inst. \\ Persistence Devices\end{tabular}} & \textbf{Ratio} \\ \hline
\rowcolor{Gray}
jiagu & 1.33M & 849K & 0.64 \\ 
smsreg & 499K & 237K & 0.48 \\ 
\rowcolor{Gray}
hiddad & 670K & 208K & 0.31 \\ 
airpush & 214K & 43K & 0.2 \\
\rowcolor{Gray}
revmob & 367K & 186K & 0.51 \\ 
dnotua & 272K & 82K & 0.3 \\ 
\rowcolor{Gray}
dowgin & 241K & 80K & 0.33 \\
leadbolt & 255K & 85K & 0.33 \\ 
\rowcolor{Gray}
mobidash & 304K & 112K & 0.37 \\ 
kuguo & 155K & 34K & 0.22 \\ 
\rowcolor{Gray}
locker & 196K & 53K & 0.27 \\ 
ewind & 139K & 17K & 0.13 \\ 
\rowcolor{Gray}
secapk & 216K & 75K & 0.35 \\ 
inmobi & 235K & 62K & 0.27 \\ 
\rowcolor{Gray}
tencentprotect & 194K & 42K & 0.22 \\ 
koler & 161K & 45K & 0.28 \\ 
\rowcolor{Gray}
domob & 176K & 45K & 0.26 \\ 
secneo & 167K & 34K & 0.21 \\ 
\rowcolor{Gray}
autoins & 209K & 33K & 0.16 \\ 
datacollector & 211K & 45K & 0.21 \\ 
\bottomrule
\end{tabular}
}
\caption{Summary of multiple-instance persistence per PHA family.}
\label{tab:pha_reinfection_top_20}
\end{table}

The fact that end users do not remove the detected PHAs promptly creates a window of opportunity (as shown in Table~\ref{tab:pha_persistence_overall}) that enables attackers to update the installed PHAs, install additional malicious apps without user consent, or entice them to install apps via full screen ads. 
We call this phenomenon \emph{multiple instance persistence}.
Figure~\ref{fig:PHA_CDF} shows the complementary cumulative distribution function (CCDF) of the number of unique PHAs installed on devices in our dataset.
A large fraction of the devices that installed PHAs installs more than one during the observation period.
For instance, 810K mobile devices (21.6\% of 3.7M devices that have at least one PHA) installed more than 7 PHAs.
In this section, we investigate to what extent the presence of a PHA on a device facilitates the installation of additional malicious components.
Our findings are shown in Table~\ref{tab:pha_reinfection_top_20}. 
18 out of the top 20 PHA families exhibit multiple-instance persistence on at least 20\% of the mobile devices they infected.
For example, 237K out of 499K mobile devices that installed PHAs from \texttt{smsreg} family have at least two other PHAs from the same family within the 14-month observation period. 
Even for \texttt{locker} and \texttt{koler} whereas the end users act swiftly (see Table~\ref{tab:pha_device_persistence_top_20}), we observe 53K (27\% of \texttt{locker} infected devices) and 45K (28\% of \texttt{koler} infected devices) exhibiting  multiple-instance persistence. 
Note that our data does not allow us to infer the causality relationship of PHA installations. 
Our results only demonstrate the fact that two PHAs from the same family \textbf{that} are installed on the same device are highly correlated.

\section{PHA In-Market Persistence}
\label{sec:pha_market}

\begin{table}
\centering
\resizebox{1\linewidth}{!} {
\begin{tabular}{ccccccc}
\toprule
\textbf{Market} & \textbf{\begin{tabular}[c]{@{}c@{}}\# Total\\ PHAs \end{tabular}} & \textbf{\#Apps}  & \textbf{\#Families}    & 
\textbf{\begin{tabular}[c]{@{}c@{}}\# Avg. Active\\ PHAs \end{tabular}} & \textbf{Active PHA (01/19 - 02/20)}\\ \midrule
\rowcolor{Gray}
\begin{tabular}[c]{@{}c@{}}Google Play\\ (com.android.vending) \end{tabular}     & 81K & 56K & 642 & 26K  & \cincludegraphics[width=1.5in]{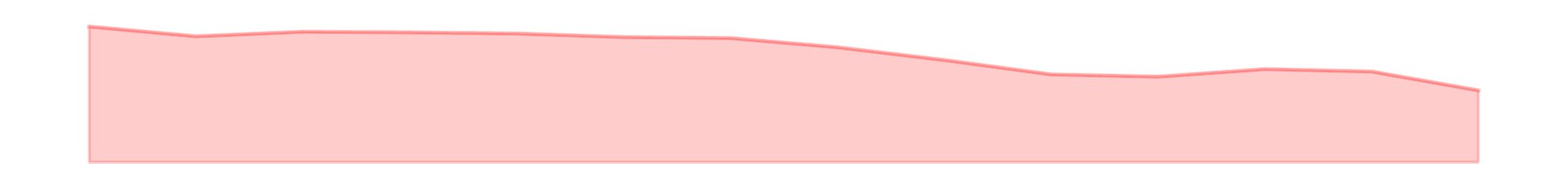}  \\ 
\begin{tabular}[c]{@{}c@{}}Huawei Market\\ (com.huawei.appmarket) \end{tabular}   & 24K & 10K  & 175 & 3K & \cincludegraphics[width=1.5in]{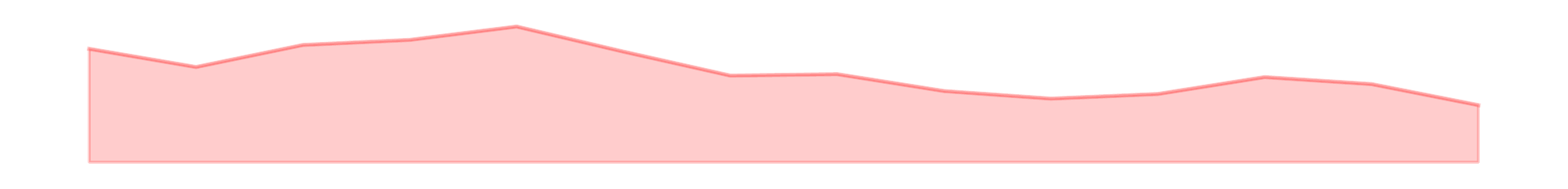}  \\ 
\rowcolor{Gray}
\begin{tabular}[c]{@{}c@{}}Xiaomi Market\\ (com.xiaomi.market) \end{tabular}    & 11K & 5K  & 226  & 2K  & \cincludegraphics[width=1.5in]{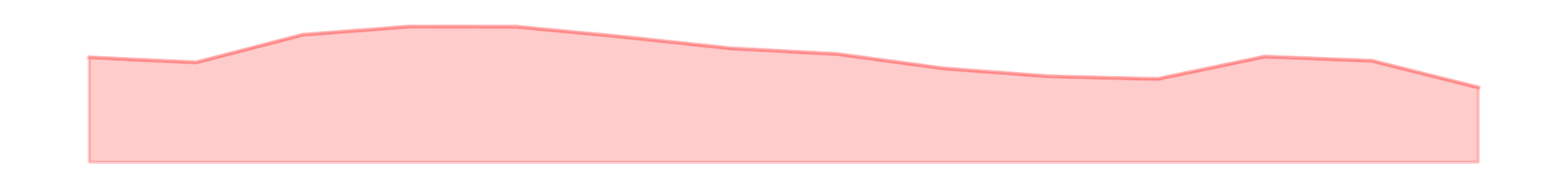}  \\ 
\begin{tabular}[c]{@{}c@{}}Samsung Market\\ (com.sec.android.app.samsungapps) \end{tabular} & 10K & 5K & 206  & 2K & \cincludegraphics[width=1.5in]{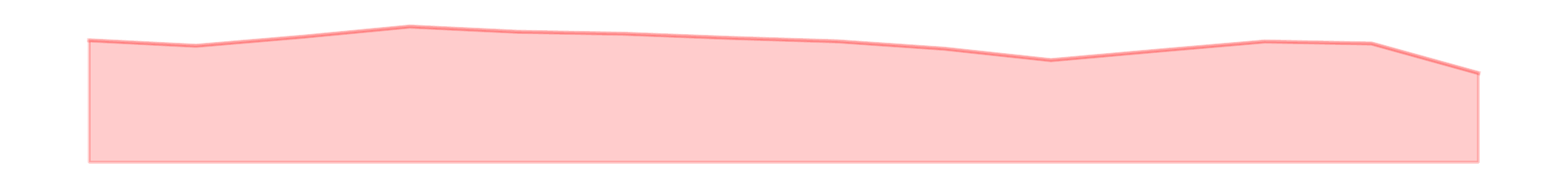}  \\
\rowcolor{Gray}
\begin{tabular}[c]{@{}c@{}}Bazaar Market\\ (com.farsitel.bazaar)  \end{tabular} & 5K & 5K & 74   & 1K & \cincludegraphics[width=1.5in]{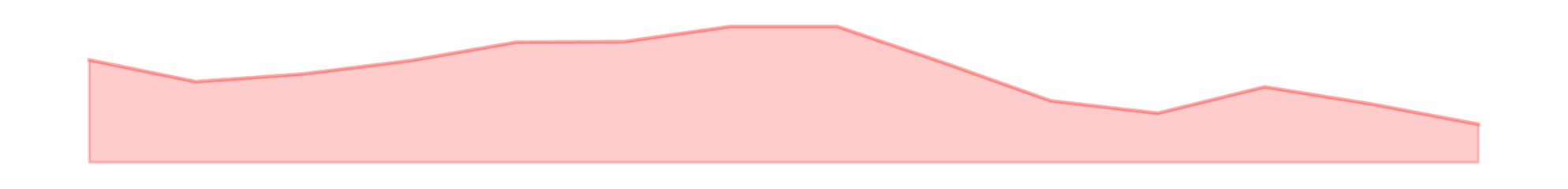}  \\
\begin{tabular}[c]{@{}c@{}}Oppo Market\\ (com.oppo.market)  \end{tabular}    & 3K & 2K & 143 & 462 & \cincludegraphics[width=1.5in]{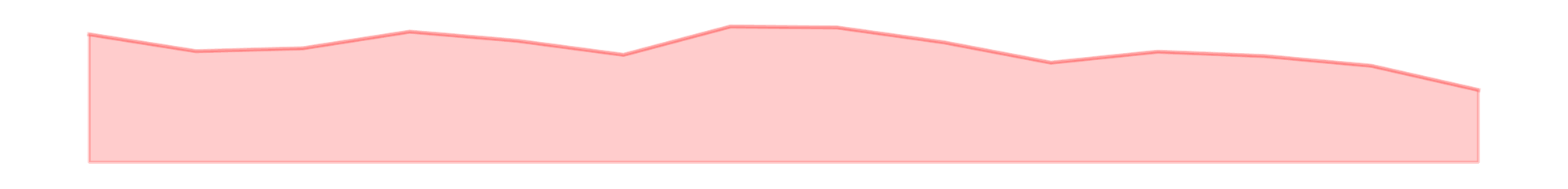}  \\ 
\bottomrule
\end{tabular}
}
\caption{Summary of active PHAs in the top 6 Android Markets.}
\label{tab:pha_active_market}
\end{table}

In this section, we first quantify the active PHAs in six app markets, by leveraging the dataset of 197K apps for which we could reliably establish their market provenance (see Section~\ref{sec:datasets}).
As discussed, this number is lower than the total number of PHAs installed for a number of reasons (\eg PHAs that were already installed on the devices before our study started, and SHA2s that do not belong to newly installed apps but are rather updates), but it still covers 66\% of all devices that installed any PHA and 22\% of all PHA installations in our dataset.
We then study how markets react to the presence of PHAs (\eg how many PHAs the markets suspend or remove, etc).
Finally, we study the PHA in-market persistence (\ie how long can PHAs persist in different markets once published) and PHA in-market evolution (\ie how PHAs may evolve to evade app vetting systems).

\subsection{PHA In-Market Prevalence} 

The mobile security product records the installer package names of PHAs observed on mobile devices (see Section~\ref{sec:datasets}). 
This enables us to track the origin market of installed PHAs.
We first measure the in-market prevalence of PHAs, serving as the foundation to understand PHA in-market persistence in Section~\ref{sec:pha_inmarket_persistence}. 
We investigate the active PHAs in six popular Android markets (\ie \texttt{Google Play}, \texttt{Huawei Market}, \texttt{Samsung Market}, \texttt{Xiaomi Market}, \texttt{Bazaar Market} and \texttt{Oppo Market}).
Our results are summarized in Table~\ref{tab:pha_active_market}. 
Google Play is the Android market hosting most PHAs: 81K unique SHA2s from 642 PHA families, with the largest monthly active PHA population (\ie 26K per month) on average. 
This makes sense, due to the fact that Google Play is the largest Android market with approximately 2.87 million apps\footnote{https://www.statista.com/statistics/266210/number-of-available-applications-in -the-google-play-store/}, and consequently becoming the de facto target of PHA makers. 
Overall, all markets demonstrate persistent monthly presence of PHAs as we can see from the temporal patterns in Table~\ref{tab:pha_active_market}. 
Following this finding, we will discuss how these markets deal with the PHAs in the next section.

\begin{table}[]
\centering

\resizebox{\linewidth}{!} {
\begin{tabular}{cccccccc}
\toprule
\textbf{Market} &  \textbf{\begin{tabular}[c]{@{}c@{}} \# Total \\ PHAs \end{tabular}} & \textbf{\begin{tabular}[c]{@{}c@{}} \# Total \\ Removed \end{tabular}}  & \textbf{\%Removed} & \textbf{\begin{tabular}[c]{@{}c@{}} \# Avg.  \\ Removed \end{tabular}}   & \textbf{\begin{tabular}[c]{@{}c@{}} PHA Removal \\ (01/2019 - 01/2020) \end{tabular}} \\ 
\midrule
\rowcolor{Gray}
Google Play    & 81K  & 74K & 91.4\% & 5.28K  &  \cincludegraphics[width=1.5in]{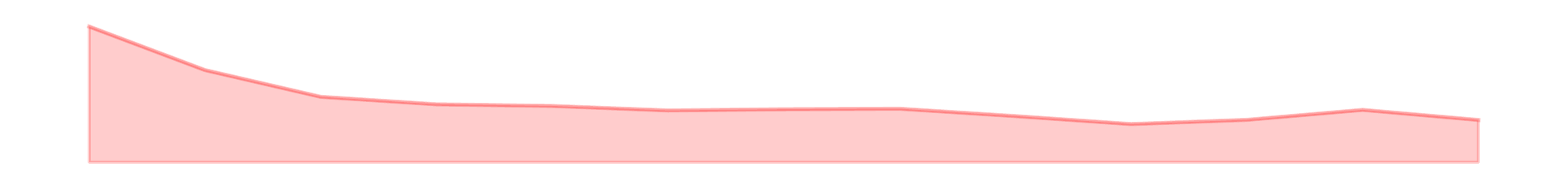} \\ 
Huawei Market  & 24K   & 22K & 91.5\% & 1.58k &    \cincludegraphics[width=1.5in]{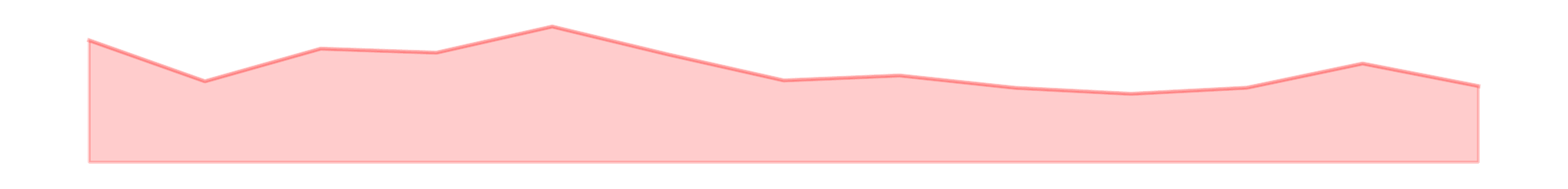}\\ 
\rowcolor{Gray}
Xiaomi Market  & 11K   & 9.8K & 92.2\% & 705 &   \cincludegraphics[width=1.5in]{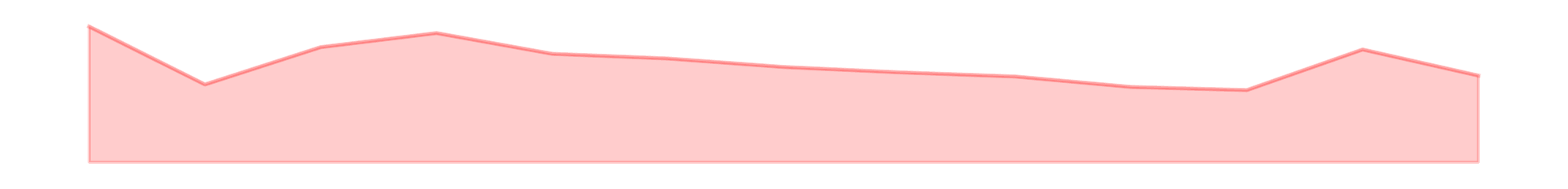}\\ 
Samsung Market & 10K & 8.9K & 91.3\% & 637  &   \cincludegraphics[width=1.5in]{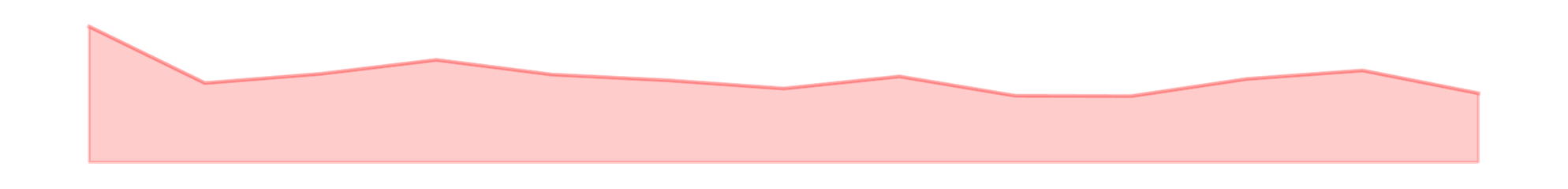}  \\ 
\rowcolor{Gray}
Bazaar Market  & 5K  & 4.7K & 95.4\% & 337 &  \cincludegraphics[width=1.5in]{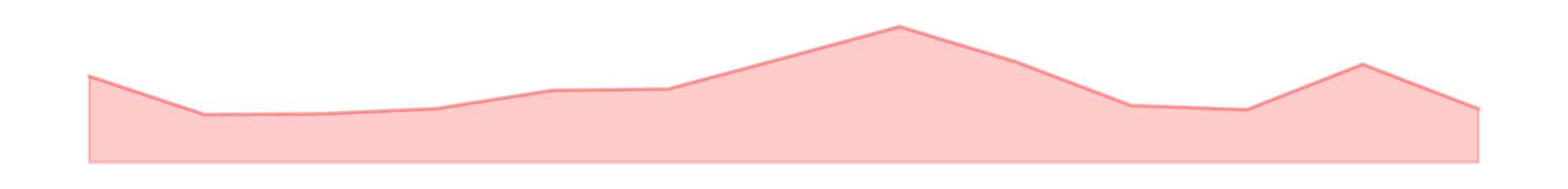}  \\ 
Oppo Market    & 3K  & 3K & 92.3\% & 223 &  \cincludegraphics[width=1.5in]{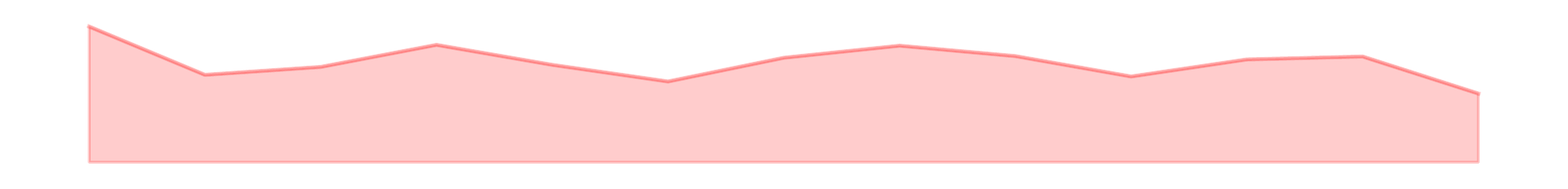}  \\ 
\bottomrule
\end{tabular}
}
\caption{Summary of PHAs removed by the top 6 Android markets.}
\label{tab:pha_removed_market}
\end{table}

\subsection{Marketplace Actions against PHAs} 
\label{sec:pha_market_response}

When apps are submitted to an Android marketplace, they are usually automatically analyzed for presence of malicious activity. 
If undetected, a PHA will be published on the marketplace, but it may later be suspended or removed, for example after the PHA is reported as malicious by users or security researchers. 
Google discloses the percentage of PHA installations in its annual Android security and privacy reports~\cite{googlereport}. 
It remains however unclear how many PHAs are suspended or removed by all marketplaces. 
As we show in the previous section, all markets demonstrate persistent monthly presence of PHAs.
For example, \texttt{Google Play} has 26K monthly active PHAs and 81K total PHAs in 14 months. 
This implies that the markets do remove PHAs, but not in a prompt manner.
To better understand this phenomenon, we follow the approach in Section~\ref{sec:approach} to measure the number of PHAs removed or suspended by the top 6 Android markets. 
Note that we define a PHA $p_i$ as removed/suspended if we do not observe the same SHA2 for the rest of our observation period after its last appearance (\ie $max(\Omega_{p_{i^{m}}})$). 
This way, we exclude all SHA2s that appeared in February 2020 to minimize false positives and discuss the limitation of this strategy in Section~\ref{sec:discussion}. 
Our findings are summarized in Table~\ref{tab:pha_removed_market}. 
Overall, each Android marketplace removes at least 91.3\% of the PHAs published on it during our observation period. 
For example, Google Play removed 74k PHAs (91.4\% of 81K PHAs) while Bazaar market performed the best removing 4.7K PHAs (95.4\% of 5K PHAs) from its market. 
Unlike previous work~\cite{wang2018beyond}, we find that Chinese marketplaces like Huawei, Xiaomi, and Oppo also remove most of the PHAs published on them (91.5\%, 92.2\%, 92.3\%). 
A reason for this discrepancy might be that our observation period is later than the ones used in previous work (2019-2020 vs 2017), and these markets might have changed their security posture after coming under scrutiny.
The temporal removal patterns of each marketplace are shown in Table~\ref{tab:pha_removed_market}, indicating that all marketplaces consistently remove PHAs. 
It is important to note that Table~\ref{tab:pha_removed_market} does not indicate that Google Play and Huawei Market are not trustworthy. 
Rather, due to the popularity of these markets and their huge user base, they consequently become the de facto targets of PHA makers. 
For instance, Google Play removed 74K PHAs during our observation period, which is respectively 8 times and 9 times more than Xiaomi Market and Samsung Market.
This is in line with the findings by Lindorfer~\etal~\cite{lindorfer2014andradar}.

\begin{table}[]
\centering
\resizebox{0.9\linewidth}{!} {
\begin{tabular}{cccc}
\toprule
\textbf{Market}   & \textbf{\begin{tabular}[c]{@{}c@{}} Average \\ Persistence \end{tabular}}  
& \textbf{\begin{tabular}[c]{@{}c@{}} Malware \\ Persistence \end{tabular}}   
& \textbf{\begin{tabular}[c]{@{}c@{}} MUwS \\ Persistence \end{tabular}}  \\ \midrule
\rowcolor{Gray}
Google Play & 77.64 D & 78.72 D & 77.11 D \\ 
Huawei Market & 30.02 D & 28.70 D & 37.61 D \\ 
\rowcolor{Gray}
Xiaomi Market & 29.93 D & 27.40 D & 37.27 D \\ 
Samsung Market & 52.56 D & 48.44 D & 81.01 D \\
\rowcolor{Gray}
Bazaar Market & 65.76 D & 65.73 D & 65.43 D \\ 
Oppo Market & 28.29 D & 26.32 D &36.47 D \\ 
\bottomrule
\end{tabular}
}
\caption{Summary of PHA in-market persistence in the Top 6 Android Markets.}
\label{tab:pha_persistence_market}
\end{table}

\begin{table*}
\centering

\resizebox{0.95\linewidth}{!} {
\begin{tabular}{cccccccccccc}
\toprule
\multicolumn{2}{c}{\begin{tabular}[c]{@{}c@{}}Google\\ Play\end{tabular}} &
  \multicolumn{2}{c}{\begin{tabular}[c]{@{}c@{}}Huawei\\ Market\end{tabular}} &
  \multicolumn{2}{c}{\begin{tabular}[c]{@{}c@{}}Xiaomi\\ Market\end{tabular}} &
  \multicolumn{2}{c}{\begin{tabular}[c]{@{}c@{}}Samsung\\ Market\end{tabular}} &
  \multicolumn{2}{c}{\begin{tabular}[c]{@{}c@{}}Bazaar\\ Market\end{tabular}} &
  \multicolumn{2}{c}{\begin{tabular}[c]{@{}c@{}}Oppo\\ Market\end{tabular}} \\ \cmidrule(lr){1-2} \cmidrule(lr){3-4} \cmidrule(lr){5-6} \cmidrule(lr){7-8} \cmidrule(lr){9-10} \cmidrule(lr){11-12}
Family &
  \begin{tabular}[c]{@{}c@{}}Avg.\\ Persistence\end{tabular} &
  Family &
  \begin{tabular}[c]{@{}c@{}}Avg.\\ Persistence\end{tabular} &
  Family &
  \begin{tabular}[c]{@{}c@{}}Avg.\\ Persistence\end{tabular} &
  Family &
  \begin{tabular}[c]{@{}c@{}}Avg.\\ Persistence\end{tabular} &
  Family &
  \begin{tabular}[c]{@{}c@{}}Avg.\\ Persistence\end{tabular} &
  Family &
  \begin{tabular}[c]{@{}c@{}}Avg.\\ Persistence\end{tabular} \\ 
\rowcolor{Gray}
airpush       & 153.3 D & jiagu          & 32.3 D & jiagu  & 30.6 D & jiagu & 39.9 D                 & adpush & 92.2 D  & \textbf{jiagu}  & 26.2 D  \\ 
jiagu         & 153.5 D & smsreg         & 46.9 D & smsreg  & 32.1 D & airpush & 188.2 D             & hiddad  & 98.7 D  & hiddad & 91.9 D  \\ 
\rowcolor{Gray}
revmob        & 159.1 D & tencentprotect & 44.3 D & umpay  & 100.1 D & revmob  & 191.8 D             & toofan & 83.5 D  & smsreg & 65.4 D  \\ 
leadbolt      & 159.9 D & secneo  &  51.8 D     & datacollector & 58.2 D  & leadbolt & 173.37D       & \textbf{privacyrisk} & 65.4  & datacollector & 46.3 D  \\

\rowcolor{Gray}
inmobi        & 125.9 D & datacollector   & 32.8 D   & tencentprotect  & 53.2 D  & smsreg  &  56.2 D  & ewind  & 129.0 D  & tencentprotect  & 52.4 D \\ 
anydown       & 191.6 D & autoins   &  41.2 D   & secneo & 31.9 D & mobby  & 194.0 D      & dnotua & 92.9 D  & utilcode & 63.0 D  \\ 
\rowcolor{Gray}
hiddad        & 165.9 D & \textbf{utilcode}    & 26.0 D  &  hiddad & 82.2 D  & tencentprotect  & 56.44 D      & hiddenapp & 80.4 D & badiduprotect & 39.3 D  \\ 
plankton      & 136.1 D & baiduprotect   &  83.5 D     & utilcode  &  52.5 D &  anydown & 183.7 D       & hiddapp & 99.7 D & beitaad & 45.9 D  \\
\rowcolor{Gray} 
datacollector & 152.2 D & autoinst    & 35.1 D  & baiduprotect & 48.9 D & \textbf{wapron} & 8.4 D   & notifyer & 75.2 D & airpush & 47.5 D  \\ 
dnotua        & 115.2 D & smspay     & 62.4 D   & \textbf{wapron}  & 13.5 D  & baiduprotect &  84.9 D    & airpush & 123.8 D & revmob & 106.6 D  \\ 
\bottomrule
\end{tabular}
}
\caption{Summary of the top 10 families (ranked by the number of SHA2s) in-market persistence in the top 6 Android marketplaces. A family name is in bold if its in-market persistence period is below average (see Table~\ref{tab:pha_persistence_market}).}
\label{tab:pha_removal_top_families}
\end{table*}

\subsection{In-Market Persistence of Different Types of PHAs} 
\label{sec:pha_inmarket_persistence}

\begin{figure}[t]
\centering
\includegraphics[width=0.7\linewidth]{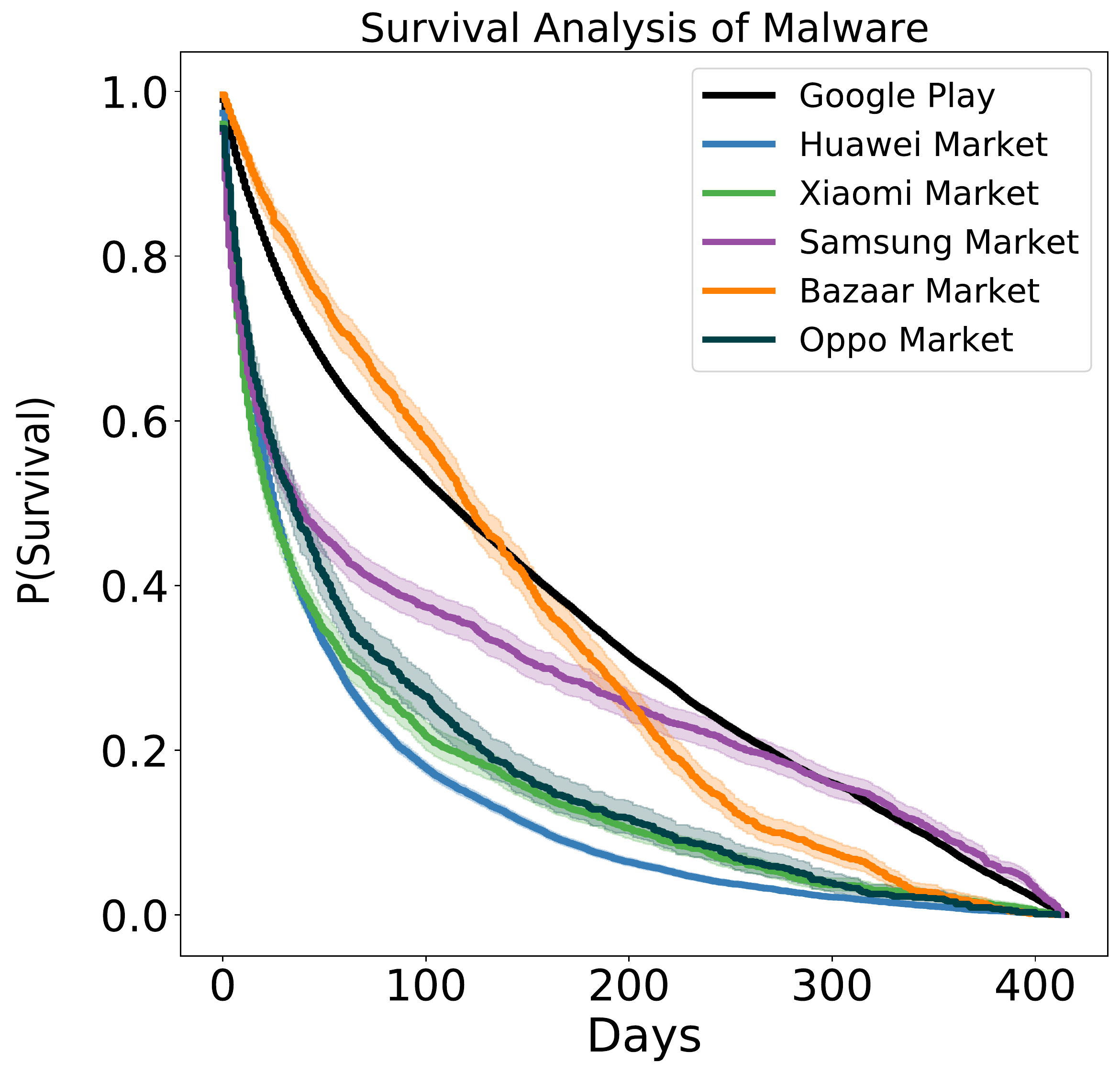}
\caption{Survival analysis of PHAs, malware and MUwS in the six markets.}
\label{fig:survival_analysis_markets}
\end{figure}

An important yet unanswered question is how long PHAs can persist in different markets before being taken down, since the longer they persist the more devices may be infected. 
To answer this question, we follow the approach outlined in Section~\ref{sec:censored} to measure PHA in-market persistence in these Android markets. 
Our findings are summarized in Table~\ref{tab:pha_persistence_market}. 
We observe that PHAs, on average, can persist in \texttt{Google Play} for 77.64 days and on other markets for at least 24 days. 
This leaves a large window of opportunity for miscreants to exploit mobile devices putting the users and their data at risk.
To further investigate the significance of our findings on mean PHA in-market persistence, we use the Kaplan-Meier Estimate~\cite{kleinbaum2010survival}. 
Recall that our methodology allows us to include censored data (see Section~\ref{sec:approach}), hence our estimates is not biased nor under-estimated.
The survival distributions of PHAs in the six markets are shown in Figure~\ref{fig:survival_analysis_markets}. 
It is visually evident that, at any point across the timeline, we can see that the survival probability of the PHAs in Google Play is more than the other markets (except Bazaar Market).
We further carry out the pairwise Peto-Prentice test to compare the survival distributions of the PHAs between Google and the other markets to establish the fact that the PHAs in Google Play persist longer than those in the other markets. 
The degrees of freedom are the number of groups minus one, hence always 1 in our tests.
A $\chi^2$ test shows that these differences are statistically significant as the test statistic values are significantly larger than 3.841 (from standard $\chi^2$ distribution table) and the $p$-values are all less than 0.005. 
These results further validate our observation.
Our observation is at odds with Lindorfer~\etal~\cite{lindorfer2014andradar}.
We hypothesize two factors that may lead to our results. 
First, Android accounts for 87\% of the global smart phone market, and, inevitably, has become the de facto target for mobile malware.
In turn, some PHAs may end up on the Google Play Store despite of Google tightening Android’s security and app review.
Second, Google Play may have different policies to address PHAs (\eg it may offer a longer grace period for these PHAs to remove offending libraries/code).
Nevertheless, Google Play removed 74K PHAs during our observation period, which is far more than those removed by the other markets.

\begin{figure}[t]
\centering
\includegraphics[width=\linewidth]{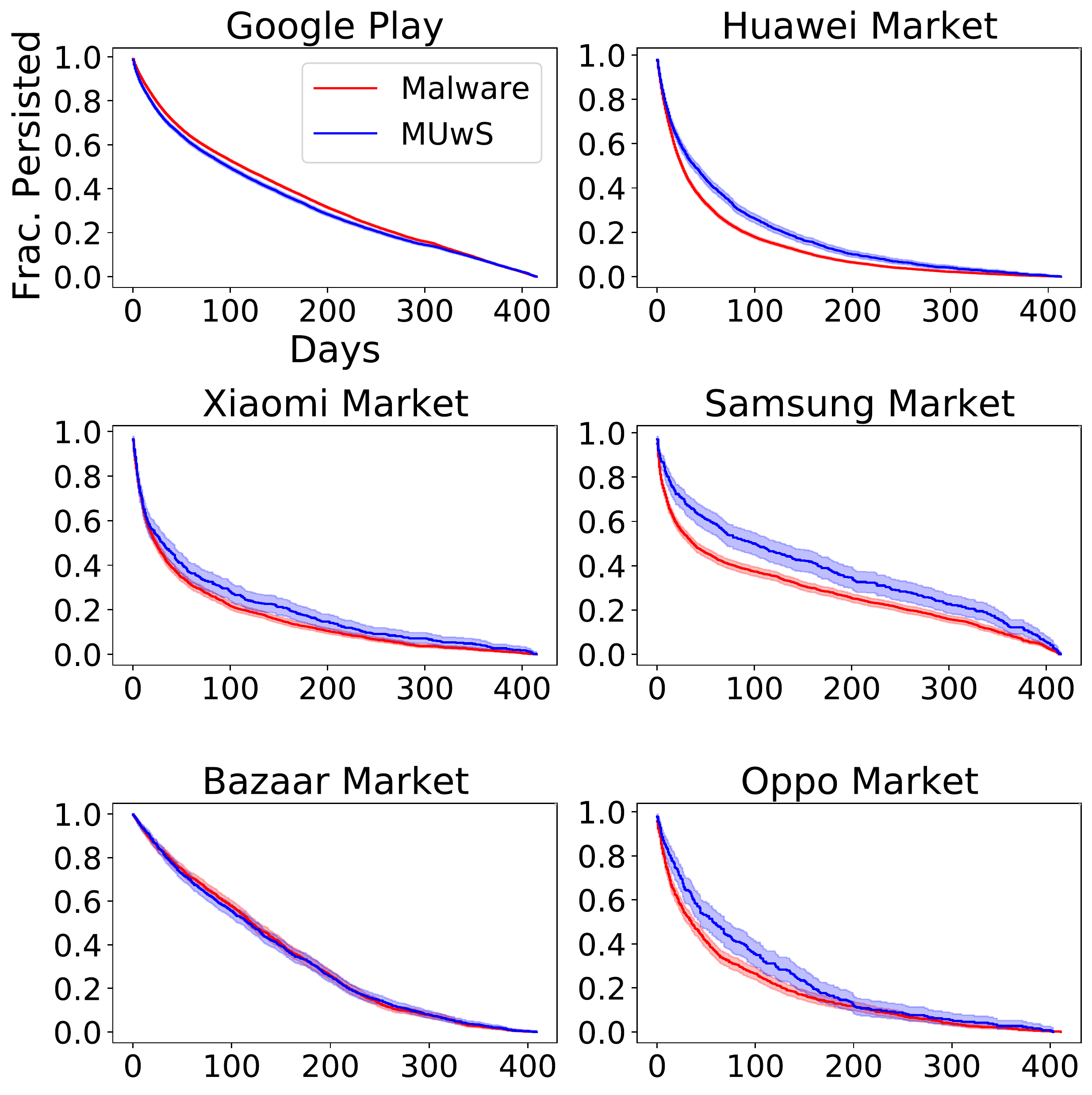}
\caption{Survival analysis of malware and MUwS persistence in the six markets.}
\label{fig:survival_analysis_malware_grey}
\end{figure}

We then use AVClass~\cite{sebastian2016avclass} to distinguish between malware and MUwS among PHAs, investigating if there exists any in-market persistence difference between these two types of PHAs.
Mobile MUwS have a slightly shorter persistence period in \texttt{Google Play} (77.11 days on average) and \texttt{Bazaar market} (65.43 days on average), while in the other four markets MUwS shows longer persistence than malware. 
In particular, the in-market persistence period of mobile MUwS is almost \emph{twice} longer than that of malware in \texttt{Samsung market} (81.01 days on average) and \texttt{Oppo market} (48.44 days on average).
This suggests that different markets apply different policies when vetting for PHAs, and might prioritize certain types of threats over others.
To further validate our findings on the in-market persistence difference between these two types of PHAs, we again use the Kaplan-Meier Estimate. 
The survival curves of mobile malware and MUwS in the six markets are shown in Figure~\ref{fig:survival_analysis_malware_grey}. 
We further carry out the Peto-Prentice test to compare the survival distributions of malware and MUwS within each market.   
A $\chi^2$ test shows that these differences are statistically significant 
as the test statistic values are significantly larger than 3.841 (from standard $\chi^2$ distribution table) and the $p$-values are all less than 0.005.
The only exception is \texttt{Bazaar market}, where the test statistic is not significant.
Hence, we cannot conclude if \texttt{Bazaar market} applies different policies when vetting for PHAs.

In Section~\ref{sec:pha_device} we showed that the overall number of devices infected is correlated with the number of SHA2s. 
Following this finding, we further study if PHA families with a large number of PHAs can persist longer in the marketplaces. 
Our hypothesis is that these large families may persist in the markets longer since app vetting systems require both machine and human inspection. 
Our findings on the top 10 largest families in the top six markets are shown in Table~\ref{tab:pha_removal_top_families}. 
We observe that most of the large families in the top six marketplaces persist longer than the mean persistence time (see Table~\ref{tab:pha_persistence_market}). 
For example, all top 10 families in \texttt{Google Play} have in-market persistence of at least 115 days, which is 38 days longer than the mean 77.64 days persistence time.      
These results show that there is a need for more comprehensive app vetting measures.

\subsection{PHA In-Market Evolution}

\begin{figure}[t]
	\centering
	\includegraphics[width=0.8\linewidth]{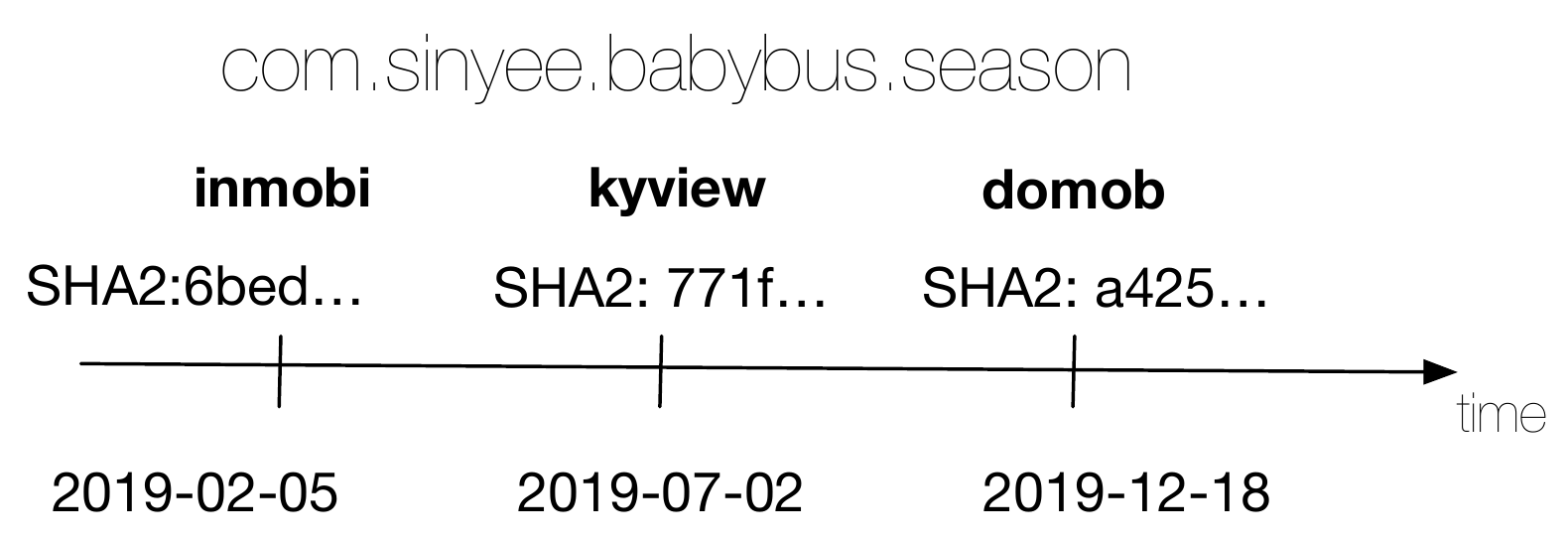}
	\caption{Example of PHA in-market evolution (\texttt{com.sinyee.babybus.season}).}
	\label{fig:pha_babybus_evolution}
\end{figure}

\begin{figure}[t]
	\centering
	\includegraphics[width=0.8\linewidth]{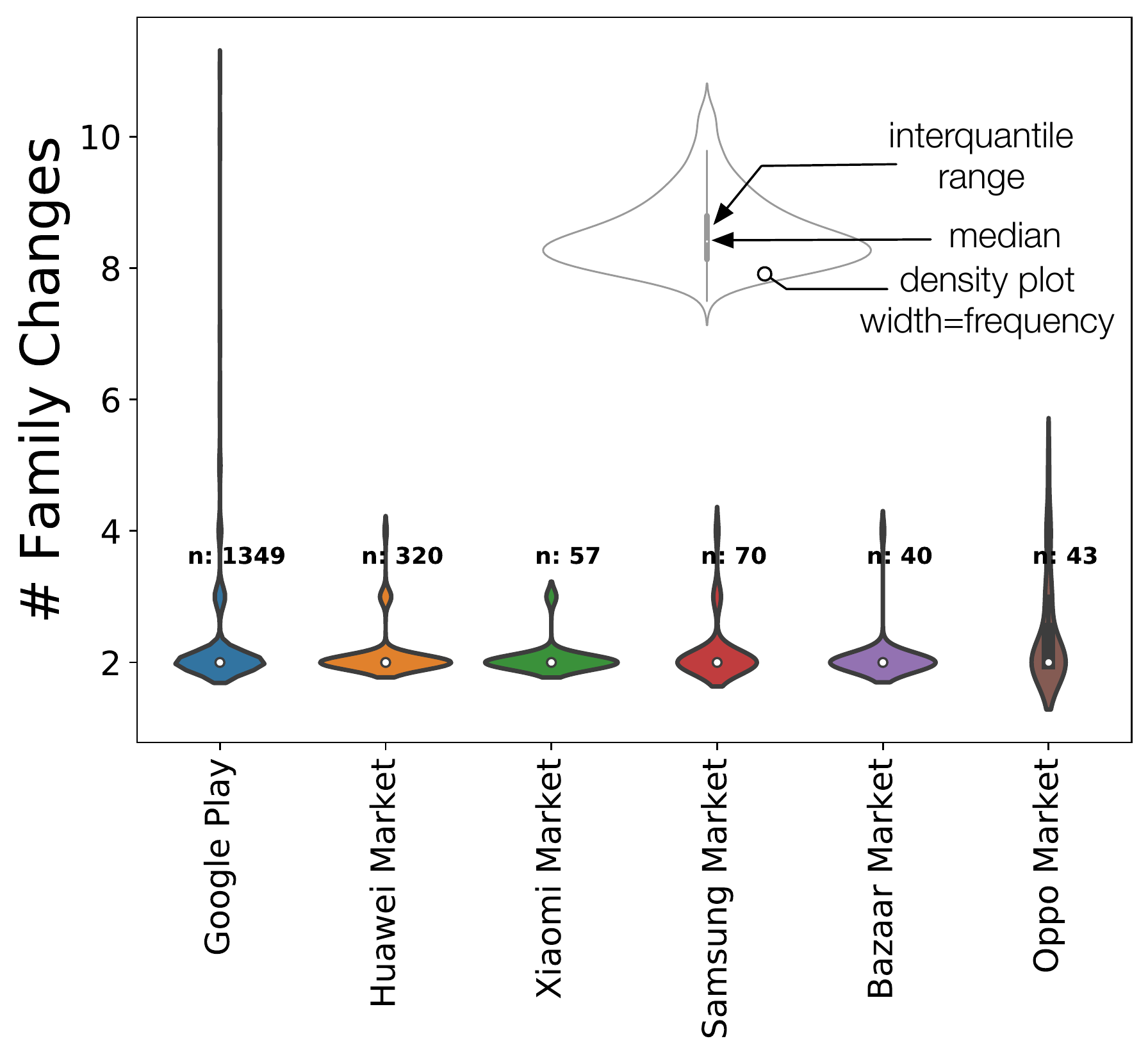}
	\caption{Violet plot summarizing PHA in-market evolution.  The white dot in the middle is the median value, the thick black bar in the centre represents the interquartile range and the contour represents the distribution shape of the data.}
	\label{fig:pha_inmarket_evolution}
\end{figure}

\begin{table}[t]
\centering
\resizebox{0.9\linewidth}{!} {	
\begin{tabular}{cccccc}
\toprule
\textbf{Market} &
  \textbf{\#Apps} &
  \textbf{\#SHA2s} &
  \textbf{\begin{tabular}[c]{@{}c@{}}Approximate\\ SHA2s\\ per PHA\end{tabular}} &
  \textbf{\begin{tabular}[c]{@{}c@{}}Avg. \\ In-market\\ Persistence\end{tabular}} &
  \textbf{\begin{tabular}[c]{@{}c@{}}Avg.\\ Evolution\\ Gap\end{tabular}} \\ \midrule
\rowcolor{Gray}
\begin{tabular}[c]{@{}c@{}}Google\\ Play\end{tabular}    & 1,349 & 7,883 & $\sim$6 & 250.1 D & 66.5 D  \\ 
\begin{tabular}[c]{@{}c@{}}Huawei\\ Market\end{tabular}  & 320   & 1779  & $\sim$5 & 276.6 D & 116.2 D \\
\rowcolor{Gray}
\begin{tabular}[c]{@{}c@{}}Xiaomi\\ Market\end{tabular}  & 89    & 443   & $\sim$5 & 247.8 D & 98.8 D  \\ 
\begin{tabular}[c]{@{}c@{}}Samsung\\ Market\end{tabular} & 70    & 383   & $\sim$5 & 238.9 D & 86.8 D  \\
\rowcolor{Gray}
\begin{tabular}[c]{@{}c@{}}Bazaar \\ Market\end{tabular} & 40    & 129   & $\sim$3 & 213.9 D & 120.5 D \\ 
\begin{tabular}[c]{@{}c@{}}Oppo\\ Market\end{tabular}    & 43    & 234   & $\sim$5 & 227.0 D & 98.4 D  \\ 
\bottomrule
\end{tabular}
}
\caption{Characteristics of PHA in-market evolution.}
\label{tab:pha_inmarket_evolution_characteristics}
\end{table}

In the previous sections, we showed that PHAs can persist in a market for weeks. 
In this section we aim to further understand how PHA families may evolve in the markets.
For example, PHA makers may proactively switch ad libraries in response to market policy changes or gain better incentives from ads, or they may modify their malicious code to evade market app vetting systems, etc.
Note that each app has a unique package name in a given market, by correlating the SHA2s belonging to a certain package name and the AVClass results of their VT reports, we can track and measure if an app evolves over time (\ie if the SHA2s of a certain PHA belong to at least 2 PHA families over the time).
We show an example in Figure~\ref{fig:pha_babybus_evolution} where SHA2s from \texttt{com.sinyee.babybus.season} in \texttt{Google Play} are associated with three different PHA families (\ie \texttt{inmobi}, \texttt{kyview}~\cite{wei2017deep}, and \texttt{domob}) during our observation period.\footnote{Note that \texttt{inmobi} is Google's preferred ad SDK partner. However, this library is flagged by multiple mobile security products as MUwS, and has leaked sensitive user data in the past. In fact, inMobi was charged by the FTC for COPPA violations in 2016. Therefore we flag \texttt{inmobi} as PHA in this paper even though we acknowledge that the definition of MUwS varies by platforms.}
Their overall evolution distribution is illustrated in Figure~\ref{fig:pha_inmarket_evolution}.
As we can see, the majority of the PHAs exhibiting in-market evolution are observed in \texttt{Google Play} and \texttt{HuaWei Market} (1,340 and 320 PHAs respectively in these two markets). 
There are a limited number of PHAs in the rest of the markets exhibiting in-market evolution. 
For example, we identify 10K PHAs in \texttt{Samsung Market} (see Table~\ref{tab:pha_removed_market}), yet only 70 of them exhibit in-market evolution. 
On average, these PHAs belong to two PHA families over time.
Additional characteristics of these PHAs exhibiting in-market evolution are summarized in Table~\ref{tab:pha_inmarket_evolution_characteristics}.
Overall, these PHAs exhibiting in-market evolution show longer in-market persistence (\ie over 200 days) in the top 6 markets. 
For example, the PHAs exhibiting in-market evolution persists in \texttt{Google Play} for 250.1 days compared to the average 77.6 days (see Table~\ref{tab:pha_persistence_market}).
The average gap between PHAs switching families in \texttt{Google Play} is 66.5 days, which is more frequent than the other markets.
We believe that the shorter gap in \texttt{Google Play} is partially due to the stringent app vetting system and security policies applied by \texttt{Google Play}. 
As such, miscreants must be proactive to deal with the scrutiny from Google.

\section{PHA Migration}
\label{sec:migration}

When their PHAs are removed from a marketplace, miscreants might migrate to alternative ones to keep their operation going.
In this section, we study how PHAs migrate among markets.

\subsection{PHA Inter-Market Migration}

\begin{figure}[t]
	\centering
	\includegraphics[width=0.8\linewidth]{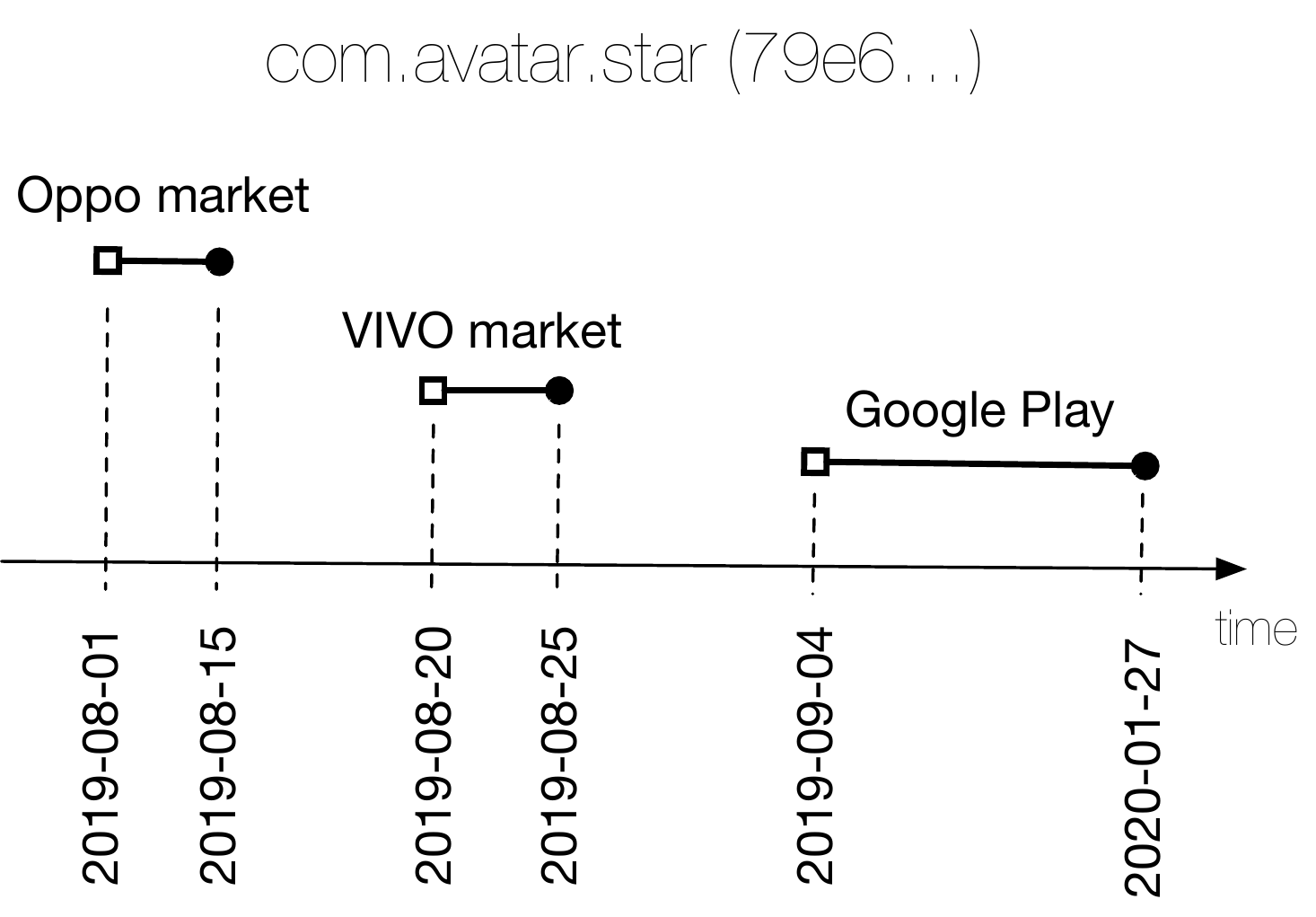}
	\caption{Example of PHA inter-market migration (\texttt{com.avatar.star}).}
	\label{fig:pha_migration_example}
\end{figure}

\begin{figure}[t]
\centering
\includegraphics[width=0.8\linewidth]{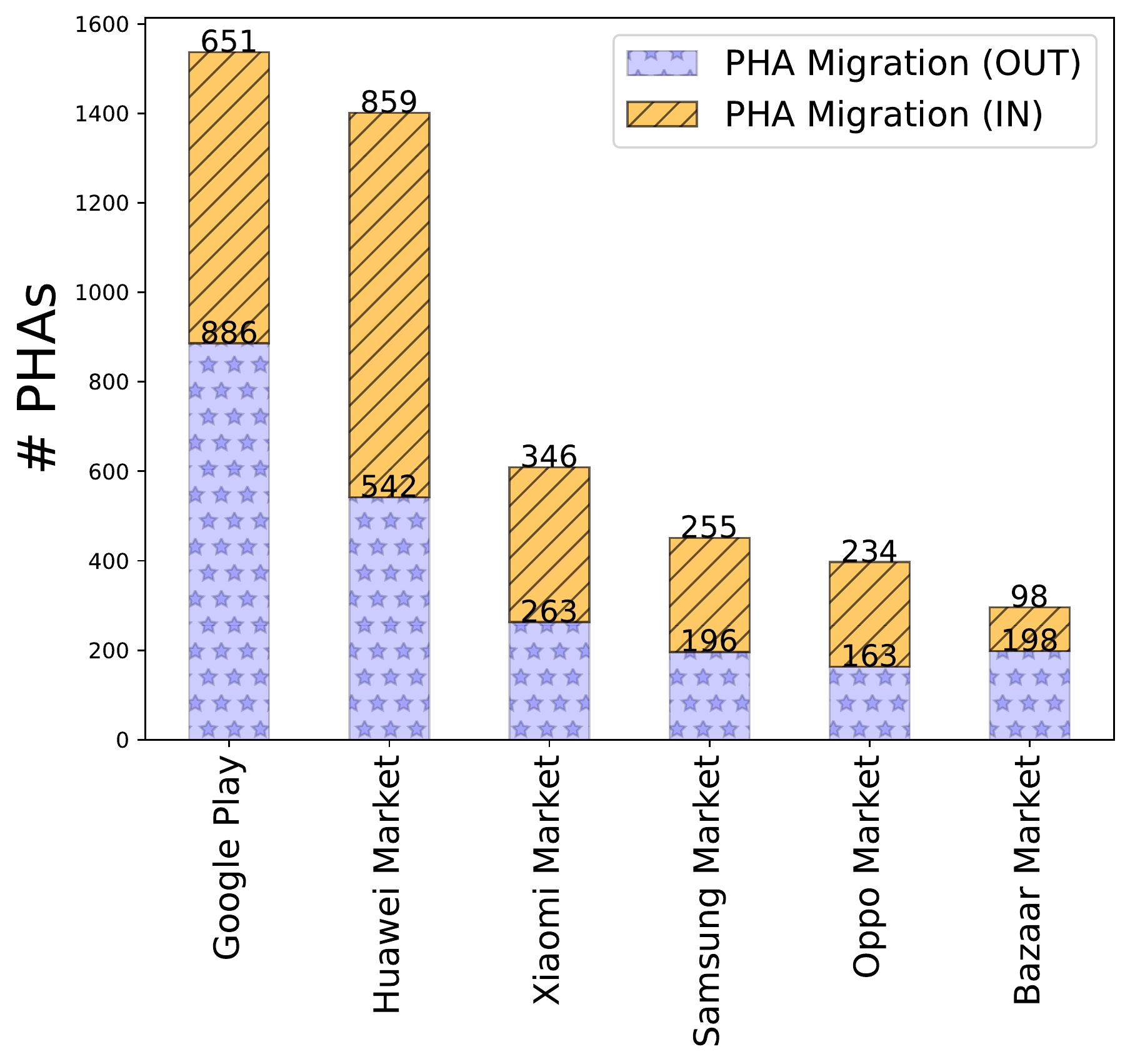}
\caption{PHA inter-market migration. Markets are ranked by the total number of inter-market activities.}
\label{tab:pha_migration}
\end{figure}

An important unanswered question is if miscreants actively move PHAs among markets to infect more devices or after such PHAs were removed from a market. 
For example, the miscreants may move a PHA to alternative markets after Google Play takes it down. 
Alternatively, the miscreants may move a PHA from alternative markets to Google Play to profit from its massive end users even for a short period of time.  
We show an example in Figure~\ref{fig:pha_migration_example}.
We have a repackaged app \texttt{com.avatar.star} with SHA2 \texttt{79e6...}, which originally appeared in \texttt{Oppo Market} between August 1, 2019 and August 15, 2019 (3 devices infected), then moved to \texttt{VIVO Market} between August 20, 2019 and August 25, 2019 (2 devices infected), and finally settled down in \texttt{Google Play} between September 4, 2019 and January 27, 2020 (215 device infected). 
Recall that we can track a SHA2 across markets and time to identify sequentially non-overlapping time intervals.
To quantify the aforementioned phenomenon, we use the app package names and their associated SHA2s as seeds (similar to previous work~\cite{lindorfer2014andradar}), and use the approach detailed in Section~\ref{sec:approach} to measure PHA inter-market migration. 
In total, we observe 3,533 PHAs that exhibit inter-market migration. 
The results for the top six markets are summarized in Figure~\ref{tab:pha_migration}. 
We see that \texttt{Google Play} exhibits the most inter-market migration activities with 1,404 PHA migrations, while \texttt{Google Play} also exhibits the most outward PHA migrations with 886 outward migration activities.

\begin{table}[t]
\centering
\resizebox{1\linewidth}{!} {
\begin{tabular}{ccccccc}
\toprule
Market &
  \begin{tabular}[c]{@{}c@{}}Total PHA\\ Migration (in)\end{tabular} &
  \# Malware &
  \# MUwS &
  \begin{tabular}[c]{@{}c@{}}Avg.\\ Persistence\end{tabular} &
  \begin{tabular}[c]{@{}c@{}}\# Device Infected \\ (upstream)\end{tabular} &
  \begin{tabular}[c]{@{}c@{}}\# Device Infected \\ (current)\end{tabular} \\ \midrule
\rowcolor{Gray}
\begin{tabular}[c]{@{}c@{}}Google \\ Play\end{tabular}  & 651 &  447 & 204 & 57.9 D & 964 & 3,003 \\ 
\begin{tabular}[c]{@{}c@{}}Huawei\\ Market\end{tabular}  & 859 & 747 & 112 &  63.5 D  & 1,039 & 1,543 \\ 
\rowcolor{Gray}
\begin{tabular}[c]{@{}c@{}}Xiaomi\\ Market\end{tabular}   & 346  & 292 & 54 & 10.44 D &  4,065 & 471  \\ 
\begin{tabular}[c]{@{}c@{}}Samsung\\ Market\end{tabular} & 255 & 218 & 37  &  23.66 D  &  1,599 & 394  \\
\rowcolor{Gray}
\begin{tabular}[c]{@{}c@{}}Oppo\\ Market\end{tabular}    &  234 & 186 & 58 & 10.3 D  &  3,364 & 296 \\ 
\begin{tabular}[c]{@{}c@{}}Bazaar\\ Market\end{tabular} & 107 & 63  & 44 & 63.84 D   &  121  & 284  \\ 
\bottomrule
\end{tabular}
}
\caption{Market response to PHA migration.}
\label{tab:market_pha_migration_in}

\end{table}

We next investigate whether mobile malware and MUwS present different migration activity on the various markets.
We again use AVClass~\cite{sebastian2016avclass} to identify mobile malware and MUwS from the package names that migrated. 
Lindorfer~\etal~\cite{lindorfer2014andradar} found initial evidence that malicious apps jump from market to market, possibly for survival.
For instance, the authors identified 131 apps that migrated to alternative markets, but didn't carry out further analysis of how long these apps would \textbf{survive} after the migration. 
To fill this gap, we then measure specifically the PHAs that migrated into the markets to understand their in-market persistence. 
The results are summarized in Table~\ref{tab:market_pha_migration_in}. 
More mobile malware migrates into the markets compared to MUwS for all top 6 markets.  
Our hypothesis is that the ecosystem of MUwS usually leverages ad libraries and can be more adaptable to market takedowns, while the miscreants behind mobile malware use more sophisticated methods (\eg code obfuscation, environment awareness, etc) hence reusing the same PHAs across different markets to maximize the number victims is more desirable.
To verify our hypothesis, we measure the device prevalence ratios of these PHAs migrating into the markets and compare this prevalence ratios to those of the immediate upstream markets they migrated from. 
Our results are summarized in Table~\ref{tab:market_pha_migration_in}.
As it can be seen, PHAs migrating into \texttt{Google Play} and \texttt{Huawei Market} (which have large user bases) manage to infect at least 50\% more devices than those from the immediate upstream markets.
However, PHAs migrating into the rest of the markets (which have smaller user bases) do not reach more devices.
Nevertheless, those PHAs, on average, have short lifespans in these markets compared to the average persistence time (see Table~\ref{tab:pha_persistence_market}, Section~\ref{sec:pha_market}) except \texttt{Huawei Market}.  
Our hypothesis is that this is partially due to the fact that these PHAs have been detected in the upstream markets, therefore signatures were made available for the downstream markets to detect them.
At the same time, the exception of \texttt{Huawei Market} shows that markets must be responsible and rigorously vet the apps submitted.  
Our study only measures the \emph{lower bound} of the PHA in-market persistence since it is possible that a PHA still exists in a market but our dataset did not reflect its existence.
The issue could be addressed if our dataset is augmented with the method proposed by Lindorfer~\etal~\cite{lindorfer2014andradar}.
We leave such task as part of our future work.

\subsection{PHA Persistence After Migration via Backup/Clone Services}
\label{sec:pha_backup_clone}

\begin{table}[t]
\centering
\resizebox{0.95\linewidth}{!} {
\begin{tabular}{cccccc}
\hline
\textbf{Service} &
  \textbf{\#PHAs} &
  \textbf{\#Malware} &
  \textbf{\#MUwS} &
  \textbf{\begin{tabular}[c]{@{}c@{}}Dev \\ Infected\end{tabular}} &
  \textbf{\begin{tabular}[c]{@{}c@{}}Avg.\\ Persistence\end{tabular}} \\ \hline
\rowcolor{Gray}
\begin{tabular}[c]{@{}c@{}} com.sec.android.easyMover \\ (Samsung) \end{tabular} & 14,038 & 10,960 & 3,078 & 35,557 & 93.38 D \\ \hline
\begin{tabular}[c]{@{}c@{}}com.samsung.android.scloud \\ (Samsung) \end{tabular} & 5,088 & 3,835 & 1,253 & 8589 & 56.41 D \\ \hline
\rowcolor{Gray}
\begin{tabular}[c]{@{}c@{}}com.hicloud.android.clone \\ (Huawei) \end{tabular} & 3,653 & 2,953 & 700 & 3,079 & 32.53 D \\ \hline
\begin{tabular}[c]{@{}c@{}}com.oneplus.backuprestore \\ (Oneplus) \end{tabular} & 1,072 & 794 & 278 & 1,361 & 22.69 D \\ \hline
\rowcolor{Gray}
\begin{tabular}[c]{@{}c@{}}com.coloros.backuprestore \\ (Oppo)\end{tabular} & 972 & 695 & 277 & 1,267 & 21.98 D \\ \hline
\begin{tabular}[c]{@{}c@{}}com.miui.cloudbackup \\ (Xiaomi) \end{tabular} & 1,243 & 928 & 315 & 1,235 & 33.23 D \\ \hline
\end{tabular}
}
\caption{PHA migration from data backup/clone services. Those services are ranked by the device prevalence ratios.}
\label{tab:pha_migration_backup_clone}
\end{table}

Android phones typically offer backup functionality to their users, allowing them to restore their apps and configuration when they purchase a new device.
This mechanism allows users to quickly restore their data (\eg contacts, settings, apps) in the new devices without manual reinstallation efforts. 
However, such services may inadvertently migrate existing PHAs to the new device too, and compromise the security and privacy of the new phones, \emph{even though these PHAs may have been removed by the markets} and therefore the user might not be able to manually install them anymore. 
Kotzias~\etal~\cite{platon2021how} showed that backup restoration is an unintended unwanted app distribution vector responsible for 4.8\% of unwanted installs. 
Following this direction, we further investigate how long PHAs can persist after migrating via backup/clone services.
Recall that the mobile security product captures an app's installer package name (see Section~\ref{sec:datasets}). 
This enables us to identify apps that were installed by backup/clone services in our dataset.
To this end, we first identify the top six data backup/clone services in our dataset and understand how many PHAs migrate from backup/clone services, and consequently how long these PHAs may persist on the devices. 
To accurately identify the data backup/clone services, we first remove all known market installer packages and rank the rest of the installers by the device prevalence ratio. 
We then investigate these apps on \texttt{Google Play} and on the Web to understand the functions of the installers.

Our findings are shown in Table~\ref{tab:pha_migration_backup_clone}. 
Overall, we observe that a considerable number of PHAs are not removed by end users and consequently  are migrated from the old phones and backups. 
For example, 14K PHAs migrated to 35.5K new Samsung models in our dataset. 
At the same time, it is interesting to see that there is three times more mobile malware than MUwS migrating via backup/clone services. 
In addition, these PHAs persist longer than the average 20.2 days persistence period (see Table~\ref{tab:pha_persistence_overall}). 
For example, PHAs migrated via Samsung smart switch mobile app (\texttt{com.sec.android.easyMover}) persist in the new devices for an averaged 93 days.

\section{\texttt{Fakeapp} Case Study}
\label{sec:case_study}

In this section, we provide a case study on the \texttt{fakeapp} Android malware family to demonstrate that our approach can measure how a malicious campaign spread, persisted, and later was removed by the markets.
\texttt{fakeapp} is a family of malicious apps that masquerades as popular legitimate apps, by using a similar package name and icons as AV apps, banking apps, etc.
Some of the apps from the \texttt{fakeapp} family may engage in malicious activities such as sending/receiving premium SMS messages and downloading other apps~\cite{zhou2012dissecting,kywe2014detecting}.

\begin{figure}[t]
     \centering
     \begin{subfigure}[t]{0.45\textwidth}
         \includegraphics[width=\textwidth,valign=b]{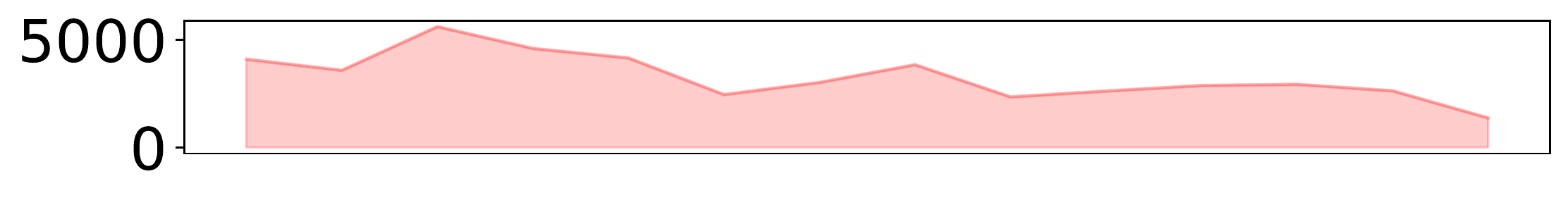}
         \caption{\texttt{fakeapp} prevalence (global)}
         \label{fig:fakeapp_case_study_global_sha2s}
     \end{subfigure}
     \hfill
     \begin{subfigure}[t]{0.45\textwidth}
         \includegraphics[width=\textwidth,valign=b]{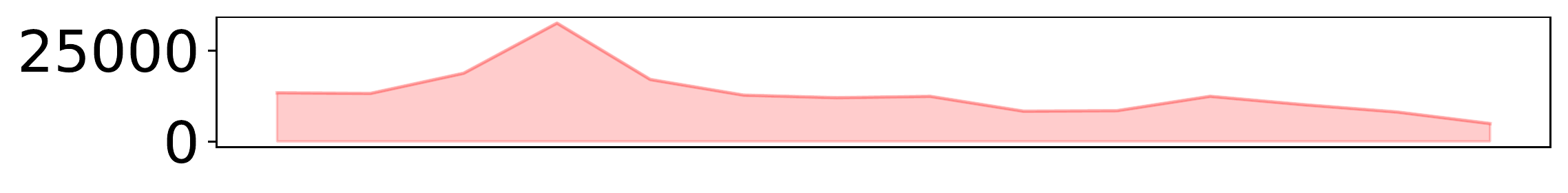}
         \caption{\texttt{fakeapp} device infection rate (global)}
         \label{fig:fakeapp_case_study_global_devices}
     \end{subfigure}
     \hfill
     \begin{subfigure}[t]{0.45\textwidth}
         \includegraphics[width=\textwidth,valign=b]{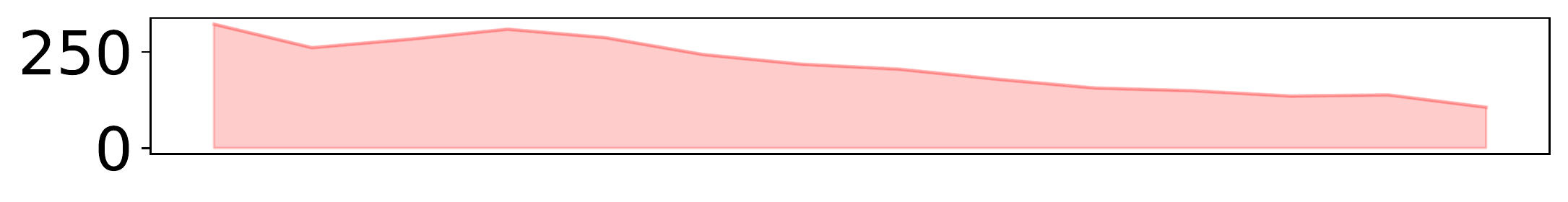}
         \caption{\texttt{fakeapp} prevalence (in Google Play market)}
         \label{fig:fakeapp_case_study_google_sha2s}
     \end{subfigure}
    \hfill
    \begin{subfigure}[t]{0.45\textwidth}
        \includegraphics[width=\textwidth,valign=b]{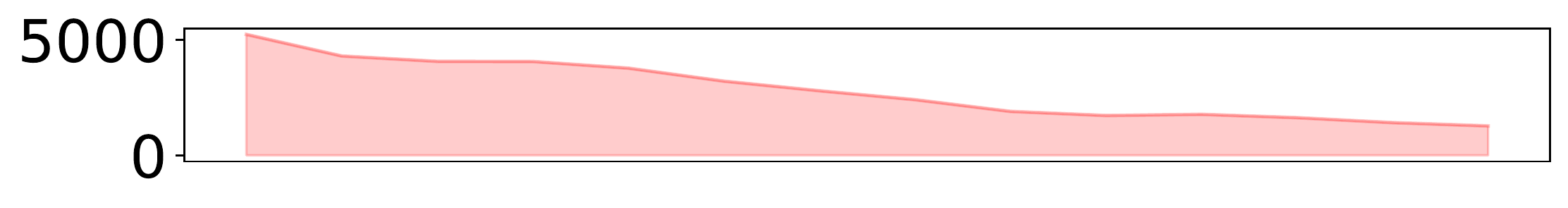}
        \caption{\texttt{fakeapp} device infection rate (incurred by Google Play)}
        \label{fig:fakeapp_case_study_google_devices}
    \end{subfigure}
    \hfill
    \caption{Case study: the \texttt{fakeapp} family}
    \label{fig:fakeapp_case_study}
\end{figure}

Figure~\ref{fig:fakeapp_case_study} shows the prevalence rate and the device infection rate of \texttt{fakeapp} from a global and a local perspective (we use the \texttt{Google Play} market in our case study). 
During our observation period, at the global level, \texttt{fakeapp} had approximately 3K active SHA2s infecting 13K devices on a monthly basis (see Figure~\ref{fig:fakeapp_case_study_global_sha2s} and~\ref{fig:fakeapp_case_study_global_devices}). 
From a local perspective, \texttt{fakeapp} had approximately 210 active SHA2s persisting in the \texttt{Google Play} market and infecting 3K devices on a monthly basis (see Figure~\ref{fig:fakeapp_case_study_google_sha2s} and~\ref{fig:fakeapp_case_study_google_devices}.
This is rather interesting because apps from the \texttt{fakeapp} family in the \texttt{Google Play} market represent 7\% of monthly active SHA2s belonging to this family, and yet accounted for approximately 25\% of the global device infection.
Our study shows that the \texttt{Google Play} market ramped up its removal efforts over our observation period.
The market had 322 active \texttt{fakeapp} SHA2s in January 2019 and reduced the number to 106 active SHA2s in February 2020, representing a twofold decrease (Figure~\ref{fig:fakeapp_case_study_google_sha2s}).
This, in turn, lead to a four-fold decrease in the number of device infection rate (Figure~\ref{fig:fakeapp_case_study_google_devices}).
However, the \texttt{fakeapp} family, on average, persisted in the \texttt{Google Play} market for 86.6 Days before removal. 
This is about 9 days longer than the average PHA persistence period on that market (see Table~\ref{tab:pha_persistence_market}, Section~\ref{sec:pha_market_response}).
At the same time, \texttt{fakeapp} apps installed from \texttt{Google Play} persist 29.41 days on devices, which is 14 days longer than the average 15.17 days \texttt{fakeapp} on-device persistence period.
We believe that this is due to the fact that \texttt{Google Play} is the \emph{de facto} trusted source of Android apps, hence the end users may keep the PHAs from \texttt{Google Play} longer. 
Additionally, the fact that \texttt{fakeapp} apps come disguised as useful apps might hide the fact that these apps are malicious and lure users into keeping them on their devices for longer, despite being warned by the mobile security product.
Our case study highlights the importance of the \texttt{Google Play} market in fighting PHAs and demonstrates the in-depth analysis that our measurement methodology can achieve.

\section{Limitations and Discussion}
\label{sec:discussion}

\mypara{Biases}
While this paper presents the largest measurement of on-device Android PHA to date, our dataset is biased towards the users of a single mobile security product, and therefore still presents some biases. 
For example, our device population is skewed towards the United States and European countries.
It is possible that end users in the United States and Europe tend to keep this mobile security app installed for longer, hence more likely that these devices fit in our data selection criteria (see Section~\ref{sec:approach}).
At the same time, we cannot observe the behavior of users that do not use mobile security products, and those who did not opt-in this data collection scheme.
Besides, we cannot observe certain events from the devices protected by Google Play Protect.
Nonetheless, we believe that our dataset is representative of the worldwide mobile users, and we do our best to minimize this bias, for example, by using percentages when looking at per country infection rates.
In terms of the representativeness of the analyzed apps, it is challenging to ascertain the coverage of our study since it is infeasible to determine the total number of all Android apps, given such a fragmented ecosystem and many alternative markets.    
Still, by analyzing 8.8M unique apps, this study is covering one of the largest sets of apps to date, and is in line with the largest datasets collected by the academic community~\cite{allix2016androzoo}.

\mypara{Data Limitations} 
It is important to note that the PHA detection data is collected passively. 
That is, a PHA detection event is recorded when the security product detects a potentially harmful application that matches a pre-defined signature including its behavior, communications, and policy violations. 
Any mobile PHAs preemptively blocked by other security products (\eg application store link blacklists, cloud-based app reputation systems) cannot be observed. 
Additionally, any PHAs that do not match the predefined signatures on devices are also not observed. 
Inferring the last seen timestamp of a PHA in a market is practically hard since the mobile security data is collected passively. Our inference therefore relies upon the deduction that if we do not observe a given PHA from billions of events generated by 11.7M devices following its last observation time, we consider that this PHA was removed by a market. 
It is possible that this PHA could still remain in that market and our dataset simply did not capture its existence (\ie this PHA is not installed by the 11.7M devices after its last observation timestamp). 
Consequently, we measure the \emph{lower bound} of the PHA in-market persistence in our study.

The Android API enables the mobile security product to identify the installer package name of a PHA.
Correlating this with the official package names of the markets, we can identify if a PHA comes from a certain market at a certain timestamp.
However, miscreants or end users can install apps via \texttt{ADB} and impersonate the official package names of the markets. 
In this case, the mobile security product can wrongly attribute a PHA as originating from a certain market.
To minimize this risk, our study only selects a PHA observed in at least two devices. 
We believe that such false positives incurred by such impersonated official market package names are statistically ignorable. 
In addition, if an app was installed before our observation period started, we cannot obtain market information for it.
If an already installer app is consequently updated, our system sees the updating software as the installer and not the original marketplace the app came from.
We therefore exclude the PHAs that we cannot confidently attribute to certain markets.
Still, this allows us to cover 66\% of the devices in our dataset and 22\% of all PHA installations.

\mypara{Implications for mobile security research}
Our study shows that many PHAs can persist on devices and in app markets for many days once installed or approved. 
We hope that our study can inspire better notification systems to nudge the end users to remove PHAs once detected, and, ideally, devise a prevention system able to convince users not to install PHAs in the first place.

\mypara{Implications to Android markets}
Our study shows that PHAs can persist in a market for at least 24 days. 
At the same time, while we recognize the efforts from the Android markets, not all PHAs are removed by them (\eg \texttt{Google Play} removes 5.28K PHAs per month and, in total, removes 74K out of 81K PHAs).
We hope that our findings will enable Android markets to ramp up their app vetting systems and takedown PHAs in a timely manner to minimize their in-market persistence.
In addition, despite of the transparency report from Google Play, we hope that the markets can be more transparent and disclose the performance figures relating to PHA removal (\eg the number of PHA removed monthly, the average time to remove a PHA, etc.).
Our study shows that PHAs may evolve over time to survive in the markets for longer and be able to reach more victims. 
We hope that our findings can encourage app markets to make end users aware of the security and privacy issues incurred by the previous versions of an app if any. 
For example, certain versions of the popular app \texttt{com.intsig.camscanner} in \texttt{Google Play} were affected by the Trojan dropper \texttt{necro} due to the integration of a 3$^{rd}$ party SDK from AdHub.
As the app remains in \texttt{Google Play} after the removal of the 3$^{rd}$ party library, a historical briefing of the security and privacy incidents associated with such apps would offer end users an informed decision when installing them on their devices in the future.

\section{Related Work}
\label{sec:related_work}

There is an enormous amount of research on mobile security and privacy.  
In this section, we specifically review previous measurement studies on malware characterization and mobile app ecosystem. 
We refer the readers to~\cite{tan2015securing,felt2011android,fang2014permission,nauman2010apex,xu2016toward,li2019rebooting} for overviews and surveys on securing Android devices.

\mypara{Mobile PHA characterization} 
The security research community has been actively investigating the ever-changing characteristics of mobile PHAs for years~\cite{tan2015securing,felt2011android,fang2014permission,nauman2010apex,xu2016toward,li2019rebooting}.
Previous efforts mainly focused on analyzing apps and systematically characterizing them from various aspects. 
From a high level, these research center on installation methods~\cite{zhou2012dissecting}, evasion mechanisms~\cite{faruki2014android}, repackaging mechanisms~\cite{zhou2012detecting,suarez2020eight,lindorfer2014andradar}, malicious payloads~\cite{zhou2012dissecting},  behaviors~\cite{lindorfer2014andrubis,yang2014droidminer}, monetization~\cite{felt2011survey}, etc. 
In recent years, Faruki~\etal\cite{faruki2014android} summarized Android security issues, malware growth (during 2010-13), their penetration, stealth techniques, and strength as well as weaknesses of some of the popular mitigation solutions. 
Mirzaei~\etal~\cite{mirzaei2019andrensemble} introduced Andrensemble, a system to characterize Android malware families by leveraging API ensembles. 
These efforts collectively shed lights on how Android malware operates in the wild, the main incentives of mobile malware, the weaknesses of some of the popular mitigation solutions, etc. 
However, they did not discuss potential threats posed by PHA persistence in both mobile devices and markets as these efforts center on app analysis and offer a less comprehensive view of the real device prevalence.

\mypara{Measurement studies on Android permission system} 
The Android permission system has been extensively covered in the previous literature~\cite{barrera2010methodology,nauman2010apex,felt2011android,au2012pscout}. We only review the work relating to our study in this paper. Felt~\etal~\cite{felt2011android} built the Stowaway system to detect overprivileged apps which could result in privacy violations. Felt~\etal~\cite{felt2012android} later showed that current Android permission warnings do not help most users make correct security decisions. Sarma~\etal~\cite{sarma2012android} discussed the risks incurred by the Android permission system and outlined 13 permissions that may critically invade users' privacy. Qu~\etal~\cite{qu2014autocog} designed AutoCog to measure the description-to-permission fidelity in Android apps and assist the end users to understand the security and privacy implications when granting permissions.

\mypara{Measurement studies on mobile PHA}  From a device perspective, Shen~\etal~\cite{shen2016insights} carried out a detailed quantitative analysis on 6.14 million Android devices comparing rooted and non-rooted Android devices across a broad range of characteristics including PHA installations and network behavior. 
Suarez-Tangil~\etal~\cite{suarez2020eight} carried out a systematic study of 1.28M repackaged apps spanning between 2010 and 2017 to understand  how Android malware has evolved over time.
More recently, Gamba~\etal~\cite{gamba2020analysis} collected 82K pre-installed apps (424K files in total) on Android devices from more than 200 vendors and carried out a measurement study to understand how the stakeholders primarily build their relationship around advertising and data-driven services.
From an app market perspective, Lindorfer~\etal~\cite{lindorfer2014andradar} proposed the AndRadar system to  discover multiple instances of a malicious Android application in a set of alternative application markets using a set of package names as seeds.
Wang~\etal~\cite{wang2018beyond} leveraged 6M Android apps downloaded from 16 Chinese app markets and Google Play and  provided a large-scale comparative study to understand various aspects and dynamics relating to apps (including PHAs), their behavior and the developers. 
These efforts collectively shed lights on the overall picture of how PHA evolves over the time.
Different from these previous efforts, our study focuses on the potential threats posed by PHA persistence in both mobile devices and markets as these efforts center on app analysis and offer a comprehensive view of the real device prevalence.

\mypara{Desktop PUP PPI ecosystem study} Another loosely connected research line is related to measuring the PUP PPI ecosystem in the PC environment. Caballero \etal~\cite{caballero2011measuring} provided the first large scale measurement of blackmarket pay-per-install services in the wild. %They achieve this by harvesting over a million client executables using vantage points spread across 15 countries. This work found that 12 out of 20 of the most prevalent malware families at the time employed PPI services to buy infections. 
Kotzias \etal~\cite{kotzias2016measuring} leveraged file dropping graphs to build a \emph{publisher graph} and identify specific roles in the ecosystem, in turn revealing the relationship between PUP prevalence and PUP distributors. Thomas \etal~\cite{thomas2016investigating} performed a similar study on unwanted software on desktop computers.

\mypara{Comparison with Close Work} 
The closest work is a recent mobile unwanted app distribution study by Kotzias~\etal~\cite{platon2021how}. 
Their study focuses on understanding who-installs-who relationships between installers and child apps, and uncovering the main unwanted app distribution vectors. 
Similar to the findings by Kotzias~\etal~\cite{platon2021how}, our study also shows that Google Play remains the main app distribution vector of PHAs, but also has the best defenses against PHAs (e.g., removing most of the PHAs).
Kotzias~\etal~\cite{platon2021how} also identifies many other distribution vectors such as bloatware, browsers, instant messaging, etc.
Our study does not cover these distribution vectors as we focus on the temporal behavior of PHAs.
Concretely, leveraging a longer observation period of PHA installation events across 11M devices, our study offers a large-scale temporal measurement study of Android PHAs to comprehend the characteristics their on-device and in-market persistence, and consequent inter-market migration after taken down. 
In summary, Kotzias~\etal~\cite{platon2021how} cover where the PHAs come from while our study addresses the temporal dynamics of PHA installations on Android.

\section{Conclusion}
We presented the largest on-device study to date of Android PHAs installed in the wild.
Our results show that PHAs on Android are a pervasive problem, and that malicious apps can persist for long periods of time both on devices and on markets.
Our results suggests that current measures against malicious apps on Android are not as effective as commonly thought, and that more research from the security community is needed in this space.

\section*{Acknowledgements}

We would like to thank our Shepherd Yousra Aafer and the anonymous reviewers for their helpful guidance through the revision process.
This work was supported by the National Science Foundation under Grant CNS-2127232.

\bibliographystyle{plain}
\bibliography{references}

\end{document}